\documentclass[runningheads]{llncs}
\usepackage[T1]{fontenc}

\usepackage{graphicx}

\usepackage{tikz}
\usepackage{multirow}
\usepackage{array}
\usepackage{xargs}
\usepackage{xspace}
\usepackage{booktabs}
\usepackage{colortbl}
\usepackage{adjustbox}
\usepackage{url}
\usepackage{soul}
\usepackage{amsmath}
\usepackage{tcolorbox}
\usepackage{float}
\usepackage{enumitem}
\usepackage[colorinlistoftodos,prependcaption,textsize=tiny]{todonotes}
\usepackage{amssymb}
\usepackage{ifthen}

\DeclareRobustCommand{\priority}[1]{\begin{tikzpicture}[scale=0.15]
    \draw (0,0) circle (1);
    \ifthenelse{#1>0}{\fill[fill opacity=0.5,fill=black] (0,0) -- (90:1) arc (90:90-#1*3.6:1) -- cycle;}{}
\end{tikzpicture}}

\newcommand{\system}{{aiXamine}\xspace}
\newcommand{\framework}{{\emph{SSP}}\xspace}
\newcommand\systemNoSpace{{aiXamine}}
\newcommandx{\yazan}[2][1=]{\todo[linecolor=red,backgroundcolor=red!25,bordercolor=red,#1]{[Yazan]: #2}}
\newcommandx{\fatih}[2][1=]{\todo[linecolor=blue,backgroundcolor=blue!25,bordercolor=blue,#1]{[Fatih]: #2}}
\newcommandx{\dorde}[2][1=]{\todo[linecolor=green,backgroundcolor=green!25,bordercolor=green,#1]{[Dorde]: #2}}
\newcommandx{\issa}[2][1=]{\todo[linecolor=orange,backgroundcolor=orange!25,bordercolor=orange,#1]{[Issa]: #2}}

\newcommand{\raid}[1]{\textcolor{black}{#1}}
\newtcolorbox{keyfinding}{
  colback=gray!5, colframe=gray!40, boxrule=0.4pt,
  sharp corners, left=6pt, right=6pt, top=3pt, bottom=3pt,
  before skip=6pt, after skip=6pt
}

\begin{document}

% \title{aiXamine: Unified Black-Box Evaluation of LLM Safety, Security, and Privacy}
\title{aiXamine: Unified Black-Box Evaluation of Cross-Dimensional Trade-offs in LLM Safety, Security, and Privacy}
\titlerunning{Cross-Dimensional Trade-offs in LLM Safety, Security, and Privacy}

\author{Fatih Deniz \and Yazan Boshmaf \and Dorde Popovic \and Issa Khalil}
\authorrunning{F. Deniz et al.}
\institute{Qatar Computing Research Institute (QCRI), HBKU, Doha, Qatar\\
\email{\{fdeniz, yboshmaf, dpopovic, ikhalil\}@hbku.edu.qa}}

\maketitle

\begin{abstract}
The critical failure modes in deployed large language models (LLMs) are cross-dimensional: a model can score 99.3 in safety alignment while refusing one in three benign queries, or improve across every capability metric while losing 21 points in privacy. Existing evaluation frameworks, which assess safety, security, and privacy independently, cannot detect these patterns. 
We introduce \system, a unified black-box platform that evaluates LLM trustworthiness across safety, security, and privacy as interdependent properties. \system orchestrates 46 tests across nine services through an automated red-teaming pipeline, producing hierarchical risk profiles, from prompt-level diagnostics to cross-service trade-off analytics, that enable reproducible comparison of proprietary and open-weight systems under identical conditions.
Applying \system to over 120 LLMs through more than 5{,}000 test runs, we conduct the largest joint safety, security, and privacy study to date and uncover three cross-dimensional phenomena invisible to single-axis evaluation. First, safety enforcement incurs a quantifiable \emph{safety tax}: stronger alignment systematically increases over-refusal, forcing providers to choose between protection and utility. Second, privacy is \raid{near-orthogonal to other trustworthiness dimensions and not captured} by standard alignment. Third, we identify and formally characterize \emph{distillation-induced robustness collapse}: off-policy distillation without on-policy correction causes entropy collapse, catastrophically destroying robustness (56.9\,$\to$\,2.6) on the same base architecture. These findings, compounded by diminishing returns from scale and category-dependent safety behaviors, demonstrate that trustworthiness is inherently multi-dimensional: progress along one axis does not guarantee, and can actively undermine, progress along others, yet current alignment methods treat it as a single objective.
\end{abstract}

\section{Introduction}
\label{sec:intro}

As LLMs are deployed in high-stakes domains such as healthcare, finance, and autonomous systems, ensuring their safety, security, and privacy has become a prerequisite for trustworthy adoption. Yet despite their exceptional capabilities, LLMs remain fragile in subtle but critical ways: they over-refuse legitimate queries~\cite{xie2024sorry}, hallucinate plausible falsehoods~\cite{ji2023survey}, succumb to adversarial manipulation~\cite{goodfellow2014explaining,kurakin2016adversarial}, amplify societal biases~\cite{mehrabi2021survey}, and leak memorized private data~\cite{fredrikson2015model,deepseek:dataleakage:blog:2025}. The community’s emphasis on capability benchmarks has left systematic trustworthiness evaluation comparatively underdeveloped~\cite{aiindex:stanford:arxiv:2024,llmethics:kumar2024ethics:arxiv:2024}, even as regulators and practitioners must navigate an ecosystem exceeding one million public models~\cite{huggingface} without standardized assessment tools.

Existing evaluation frameworks capture fragments of this risk landscape but lack the breadth to reveal cross-dimensional interactions. HELM~\cite{helm:liang2022holistic:arxiv:2022} and TrustLLM~\cite{trustllm:huang2024:2024:arxiv} omit privacy or code security; DecodingTrust~\cite{decodingtrust:wang2023decodingtrust:neurips:2023} and TrustEval~\cite{trusteval:wang:acl:2025} provide partial coverage but do not jointly assess jailbreak resistance, privacy leakage, code safety, and over-refusal within a single reproducible methodology. Most benchmarks also assume white-box access, precluding evaluation of proprietary systems under realistic deployment conditions~\cite{mohan2024securing}. Meanwhile, regulators increasingly demand transparent assurance under frameworks such as the EU AI Act~\cite{ebers2023european} and NIST AI Risk Management Framework~\cite{nistai2024artificial}. This fragmentation obscures the trade-offs and interdependencies between safety, security, and privacy, \raid{interactions that significantly influence real-world deployment risk}.

\systemNoSpace{}\footnote{Publicly available at \url{https://aixamine.qcri.org/}. Evaluation reports are accessible for reproducibility, and users may submit and run their own examinations.} addresses these gaps with a unified, black-box methodology for assessing LLM safety, security, and privacy (\framework). We integrate 46 standardized tests across nine services, spanning adversarial robustness, code security, fairness and bias, hallucination, jailbreak robustness, model and data privacy, out-of-distribution robustness, over-refusal, and safety alignment, producing structured per-model risk profiles that enable consistent comparison across proprietary and open-weight systems. We support context-dependent assessment through configurable stakeholder weightings, enabling trustworthiness evaluation aligned with deployment-specific risk priorities. Unlike capability leaderboards that emphasize accuracy, \system characterizes risk boundaries: mapping adversarial failure modes, quantifying safety\textendash utility trade-offs, and testing whether safety behaviors generalize across categories. This cross-service perspective reveals \raid{recurring patterns, including the relative independence of privacy from other trustworthiness dimensions}, that single-metric evaluations cannot observe. \raid{These observations motivate a security-centric view of trustworthiness: deployment risk is often determined not by a single failure mode, but by the interaction between multiple properties.}

\raid{Unlike broad trustworthiness leaderboards that emphasize capability comparison, \system targets security-relevant failure modes under realistic black-box deployment. Its distinctive contribution is not measuring adversarial and privacy failures in isolation, but exposing how they interact with related trustworthiness properties (over-refusal as a service-availability failure, hallucination as integrity failure) to determine deployable security posture. The distillation-induced robustness collapse we characterize illustrates how development choices can profoundly alter security posture despite preserving much of a model's apparent capability.}

\raid{Because frontier models are increasingly developed through scaling, alignment, and distillation, we investigate how these development choices affect trustworthiness outcomes.} Specifically, we investigate the following research questions:
\begin{itemize}[leftmargin=*]
\itemsep0em
\item \textbf{RQ1.} How do safety, security, and privacy interact? Is there a fundamental trade-off between safety enforcement and utility, and are these dimensions jointly addressed by current alignment?
\item \textbf{RQ2.} What drives a model’s safety profile: scale or alignment strategy? Is there a minimum model size required for safety?
\item \textbf{RQ3.} Does advancing model capability preserve privacy?
\item \textbf{RQ4.} Do safety behaviors generalize across risk categories, or are they narrow learned patterns?
\end{itemize}

Throughout this work we distinguish three complementary failure modes: \emph{safety} failures arise during intended use (hallucinations, harmful outputs, excessive refusals), \emph{security} failures arise under adversarial attack (jailbreaks, prompt manipulation, input perturbations), and \emph{privacy} failures involve inappropriate disclosure of sensitive information. Each requires distinct evaluation methodology, formalized in Section~\ref{sec:framework}.

Our evaluation of over 120 LLMs through more than 5{,}000 black-box test runs reveals patterns obscured by single-score leaderboards. We find that safety enforcement incurs a measurable utility cost, \raid{privacy is near-orthogonal to other trustworthiness dimensions}, distillation can collapse adversarial robustness from 56.9 to 2.6 on the same base architecture, \raid{in most family pairs examined, the newer generation exhibits a privacy regression} despite gains elsewhere, and safety behaviors are category-dependent rather than universal. Even the top-ranked model is not immune: GPT-5 leads the overall ranking yet loses 21.7 points in Privacy relative to its predecessor, despite gaining 3.8 points overall. These findings demonstrate that trustworthiness is inherently multi-dimensional: progress along one axis does not guarantee, and can actively undermine, progress along others. Improving it will require moving beyond monolithic alignment toward modular, dimension-aware safety architectures.
In summary, our contributions are:
\begin{itemize}[leftmargin=*]
\itemsep0em
\item \textbf{Large-Scale Cross-Model \raid{Security \& Trustworthiness} Study.} An analysis of 120+ LLMs across 5{,}000+ evaluations showing that trustworthiness is inherently multi-dimensional: safety is category-dependent, \raid{privacy is near-orthogonal to other dimensions}, and safety enforcement introduces measurable utility trade-offs.
\item \textbf{Unified Evaluation Framework.} A black-box methodology integrating 46 tests across nine services to jointly assess safety, security, and privacy under a single framework (\framework). \system further supports context-dependent assessment through configurable stakeholder weightings.
\item \textbf{Distillation-Induced Robustness Collapse.} We identify a structural vulnerability in strong-to-weak distillation pipelines: the omission of on-policy correction causes entropy collapse and adversarial brittleness, providing the first robustness-focused analysis of this failure mode.
\end{itemize}
\section{Overview}

\subsection{Threat Model}

We consider two complementary risk settings: adversarial threats from a deliberate attacker and emergent risks under benign use.

\noindent\textbf{Adversarial setting.}
The adversary operates under a black-box threat model with unrestricted query access but no knowledge of model internals, gradients, or system prompts. The adversary's goals are to (i)~elicit unsafe or policy-violating outputs, (ii)~extract memorized training data or personally identifiable information, (iii)~bypass alignment safeguards and refusal mechanisms, or (iv)~induce generation of insecure code. An attack succeeds if the model produces any output that violates the targeted safety, security, or privacy objective as determined by our scoring pipeline.

\noindent\textbf{Benign-use setting.}
Even absent adversarial intent, models may exhibit emergent failures that compromise trustworthiness under normal operation, including hallucinating facts, producing unfair or biased outputs, generating vulnerable code, or refusing harmless queries.

\noindent\textbf{Evaluation constraints.}
All evaluations comply with provider rate limits and leave content-moderation filters enabled: we evaluate systems as deployed, because that is what users encounter. This ensures uniform applicability across proprietary and open-weight deployments.

\subsection{Evaluation Platform}
\label{subsec:eval_framework}

\begin{figure}
    \centering
    \includegraphics[width=0.8\columnwidth]{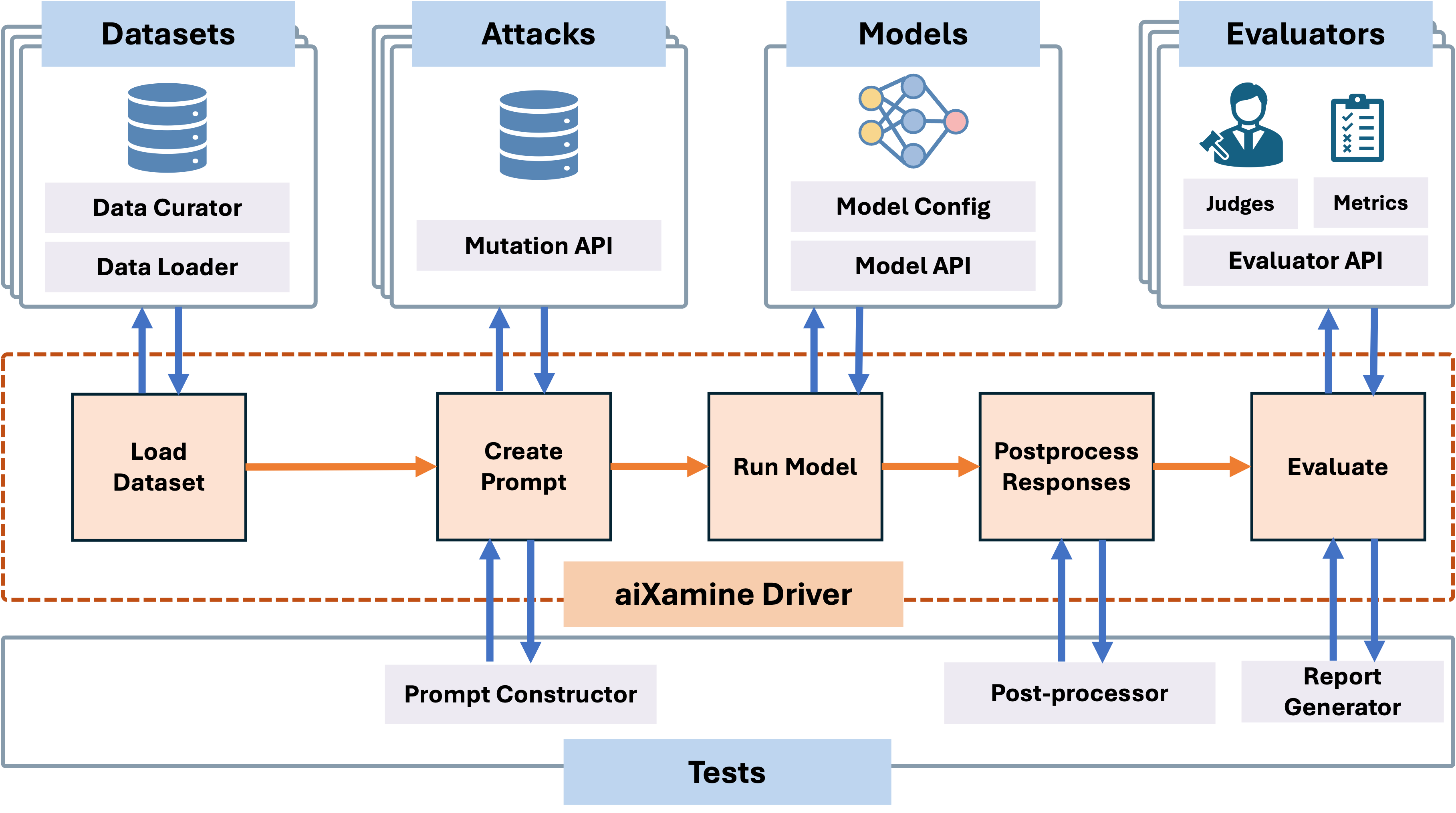}
\caption{\system’s evaluation pipeline. Benchmark prompts and their adversarial and OOD variants are dispatched under standardized inference parameters; responses are scored by deterministic rules or multi-judge ensembles; and normalized scores are aggregated at the service, category, and subcategory levels.}
    \label{fig:overview}
\end{figure}

Figure~\ref{fig:overview} depicts the \system pipeline. Benchmark prompts, including adversarial and out-of-distribution variants, are dispatched to each registered model under standardized inference parameters. Responses are scored by deterministic rules or multi-judge ensembles, and normalized scores are aggregated hierarchically at the service, category, and subcategory levels. All datasets, configurations, and scoring scripts are publicly released for reproducibility. Table~\ref{tab:datasets} in Appendix~\ref{appendix:overview} enumerates the complete test suite, listing each service, dataset, sample count, and metric.

Evaluations follow a four-level taxonomy (services, tests, categories, and subcategories) structured around the \framework framework (Figure~\ref{fig:framework}), which decomposes trustworthiness into three complementary dimensions: \emph{Safety \& Reliability} (factual consistency, ethical alignment, refusal behavior), \emph{Security \& Robustness} (resilience against adversarial inputs, jailbreaks, and distribution shifts), and \emph{Privacy \& Fairness} (confidentiality, impartiality, demographic equity). These dimensions are operationalized through nine services: \textit{Adversarial Robustness}, \textit{Code Security}, \textit{Fairness \& Bias}, \textit{Hallucination}, \textit{Jailbreak Robustness}, \textit{Model \& Data Privacy}, \textit{Out-of-Distribution Robustness}, \textit{Over Refusal}, and \textit{Safety Alignment}. Each service is instantiated through standardized datasets, domain-specific scoring logic, and judge ensembles. Tests within a service probe specific risk types (semantic transformations for adversarial stress-testing, demographic pairing for fairness, structured extraction for privacy leakage, grounding checks for hallucination), and responses are decomposed into category and subcategory scores that localize weaknesses (e.g., excessive refusals in health-related prompts, memorization of named entities, or unsafe completions in specific programming languages). Detailed per-service definitions are provided in Section~\ref{sec:framework}.

\subsection{Evaluation Metrics}

A core design requirement is that all metrics reside on a common $[0,1]$ scale so that scores from heterogeneous services can be aggregated and compared without unit-dependent distortions. Each test produces three analytical layers: atomic scores for individual prompts, service-level aggregates within a defined competency, and cross-service analytics that quantify interdependencies between dimensions.

The primary metric is normalized accuracy: the fraction of responses satisfying the test criterion (e.g., refusing unsafe instructions under \textit{Safety Alignment}). Where accuracy is inapplicable, we use task-specific alternatives: harmful-content pass-rate, privacy-leakage rate, refusal appropriateness, and faithfulness score. Service-level scores are the arithmetic mean of their constituent test scores.

Non-standard metrics require principled normalization to remain interpretable alongside accuracy. For fairness, we adopt Cram\'er's $V$ rather than the $\chi^2$ statistic: $\chi^2$ is sample-size dependent and yields a test statistic whose magnitude is not directly comparable across datasets, whereas $V$ normalizes $\chi^2$ to $[0,1]$, producing an effect-size measure that can be read on the same scale as accuracy. Because stronger associations indicate greater disparity, we report $1 - V$ so that higher values consistently reflect lower bias. Similarly, Pearson’s $r$ (used for agreement between model and human judgments, e.g., privacy awareness) is linearly rescaled from $[-1,1]$ to $[0,1]$ after direction alignment, preserving its role as an agreement-quality indicator within the unified scoring framework.

\subsection{Evaluation Challenges}

Beyond the fragmentation and limited cross-dimensional coverage of existing benchmarks discussed in Section~\ref{sec:intro}, several methodological challenges require dedicated solutions.

\noindent\textbf{Inconsistent scoring and judge calibration.}
Evaluations rely on automated judges whose reliability varies across datasets. LLM-based evaluators misjudge edge cases or abstain, and studies use incompatible metrics (accuracy, refusal rates, moderation scores), hindering reproducibility. We mitigate this through a multi-judge ensemble combining open and closed-weight models, rule-based evaluators, and external moderation APIs. To validate that our results are not artifacts of a single judge, we independently scored the full evaluation suite with two general-purpose judges and confirm near-perfect agreement across all service-level scores.

\noindent\textbf{Instruction non-compliance.}
Models frequently deviate from evaluation instructions, refusing benign queries or ignoring format directives, distorting measured safety. Benchmarks rarely disclose how non-compliance is detected or corrected. We address this with separate refusal and compliance judges that differentiate genuine safety refusals from generic abstentions, improving interpretability and fairness in scoring.

\noindent\textbf{Ranking stability under aggregation.}
A single aggregate ranking implicitly assumes uniform importance across all dimensions, yet a healthcare deployment and a code assistant have fundamentally different risk priorities. We validate robustness by reranking under six stakeholder profiles and 10{,}000 Dirichlet-sampled weight vectors, confirming that coarse tier membership is robust while within-tier ordering is sensitive to the evaluation perspective.

These challenges, together with the fragmentation and coverage limitations discussed earlier, motivate the design principles formalized in Section~\ref{sec:framework}.
\section{The \framework Framework}
\label{sec:framework}

\begin{figure}
    \centering
    \includegraphics[width=0.75\columnwidth]{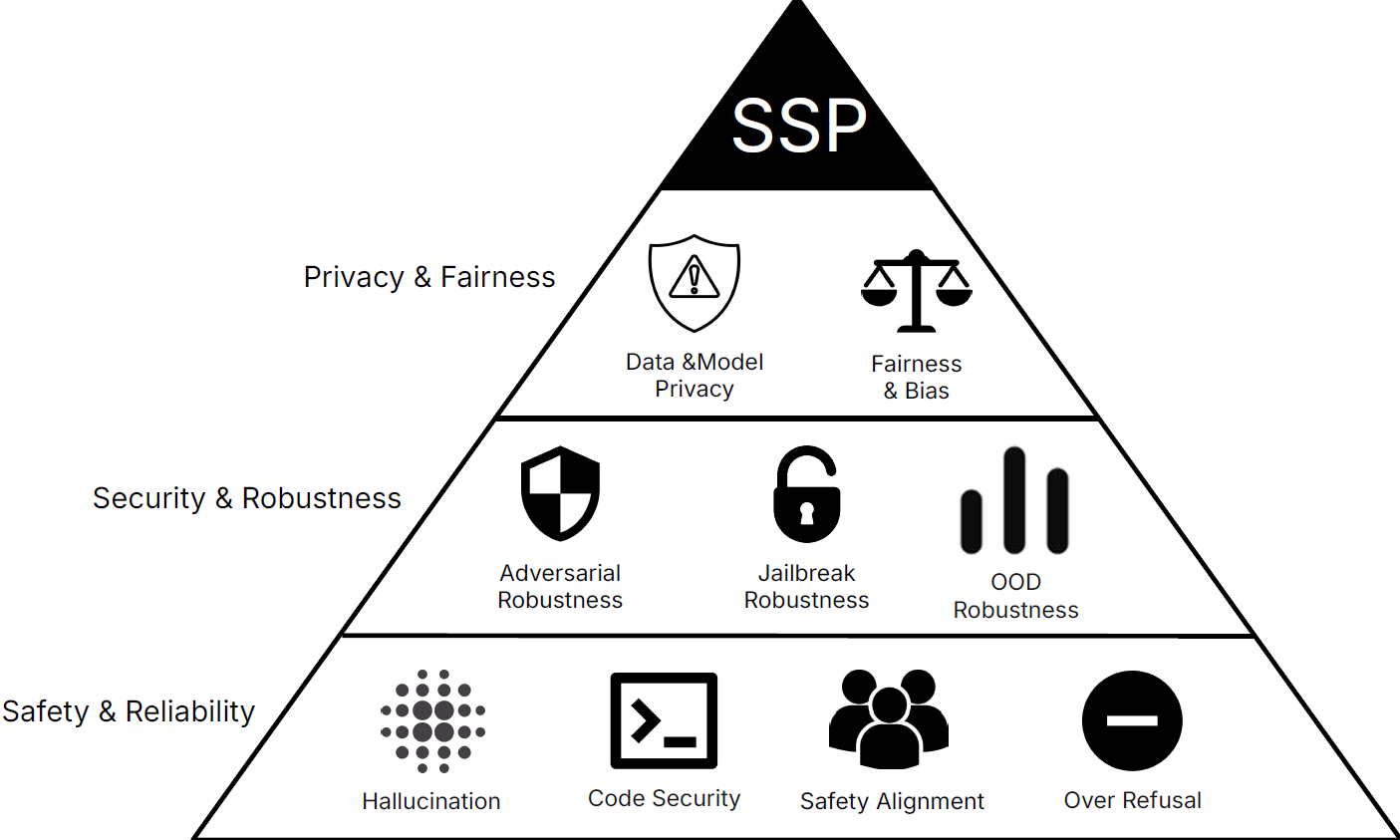}
\caption{Hierarchical structure of the SSP evaluation framework. The three dimensions form complementary levels of trustworthiness: Safety \& Reliability for behavioral integrity, Security \& Robustness for adversarial resilience, and Privacy \& Fairness for societal trust. Each level aggregates multiple services, each comprising test suites that evaluate core competencies in model safety, security, and privacy.}
    \label{fig:framework}
\end{figure}

While Safety, Security, and Privacy are often conflated, they reflect distinct failure modes. \textit{Safety} concerns harmful behavior during intended use (e.g., hallucinations, ethical violations, over-refusal). \textit{Security} captures adversarial compromise, including jailbreaks, prompt manipulation, and code vulnerabilities. \textit{Privacy} addresses the exposure of sensitive information, from memorization-based leakage to improper disclosure at inference time. \framework organizes these into a three-level hierarchy (Figure~\ref{fig:framework}), where trustworthiness progresses from internal reliability, to robustness under stress, to societal assurance. Each level builds on the previous, with lower-level failures undermining higher-level guarantees. \framework is operationalized through nine services described below.

\raid{The nine services composing SSP were synthesized from trustworthiness dimensions commonly evaluated in prior frameworks, including HELM~\cite{helm:liang2022holistic:arxiv:2022}, TrustLLM~\cite{trustllm:huang2024:2024:arxiv}, DecodingTrust~\cite{decodingtrust:wang2023decodingtrust:neurips:2023}, and TrustEval~\cite{trusteval:wang:acl:2025}. Table~\ref{tab:coverage-by-service} in Section~\ref{sec:relatedwork} maps the dimensions covered by these frameworks to the SSP services. While prior frameworks typically present these dimensions as largely independent categories, SSP organizes them into a three-layer hierarchy according to the aspect of trustworthiness they address: behavioral integrity (Safety \& Reliability), adversarial resilience (Security \& Robustness), and societal trust (Privacy \& Fairness).}

\subsection{Safety \& Reliability}
The base layer focuses on the internal behavioral integrity of LLMs, determining whether a model behaves predictably and responsibly in potentially sensitive contexts.

We evaluate model behaviors across these contexts through four services: \textit{Hallucination}, which measures factual accuracy and grounding; \textit{Code Security}, which assesses whether generated code is free from vulnerabilities and unsafe patterns; \textit{Safety Alignment}, which tests adherence to ethical and safety guidelines; and \textit{Over Refusal}, which assesses unnecessary conservatism in safe interactions.

\noindent\textbf{Hallucination.} This service evaluates whether a model produces factual, self-consistent content grounded in its provided context. We employ seven complementary tests to achieve this: \textit{SimpleQA}~\cite{simpleqa:arxiv:wei2024measuring:2024}, \textit{TruthfulQA}~\cite{truthfulqa:lin2021truthfulqa:arxiv}, and \textit{TriviaQA}~\cite{dataset:triviaqa:joshi:2017} to examine factual accuracy and resistance to common human misconceptions in open-domain question answering; \textit{SelfCheckGPT}~\cite{selfcheckgpt:arxiv:manakul2023selfcheckgpt:2023} to measure internal consistency across repeated generations, revealing self contradictions or unstable reasoning; \textit{Vectara}~\cite{vectara:leaderboard:2024} and \textit{FaithEval}~\cite{faitheval:shafiq:2024} to assess contextual faithfulness in summarization and retrieval-augmented generation tasks; and \textit{HaluEval}~\cite{halueval:arxiv:li2023halueval:2023} to extend the analysis to multi-turn dialogue and reasoning, evaluating whether conversational outputs remain coherent and supported by evidence.

\noindent\textbf{Code Security.} This service evaluates whether a model generates code that is functionally correct, free from exploitable vulnerabilities, and consistent with secure development practices. We employ two complementary tests, \textit{CyberSecEval 3}~\cite{code:arxiv:wan2024cyberseceval:2024} and \textit{SecCodePLT}~\cite{code:arxiv:yang2024seccodeplt:2024}, each in two variants: instruction-based, where the model writes code from scratch, and autocomplete, where it completes partial code that may include unsafe priors. \textit{CyberSecEval 3} examines vulnerability avoidance across six programming languages and 50 CWE categories in both settings, revealing whether models implicitly apply secure-coding principles. \textit{SecCodePLT} targets Python-specific weaknesses such as command injection, unsafe deserialization, and improper resource handling, assessing fine-grained reasoning about security in concrete programming contexts.

\noindent\textbf{Safety Alignment.} This service assesses whether a model consistently recognizes and mitigates harmful, unethical, or policy-violating content across diverse risk categories (e.g, harmful language, misinformation, and ethics, and others in WildGuard test). We integrate extensive safety datasets~\cite{anthropicredteaming:bai2022training:arxiv:2022,dataset:beavertails,dataset:harmfulqa,dataset:donotanswer,openai2022moderation,dataset:realtoxicityprompts,dataset:simplesafetytests:vidgen2023:arxiv:2023,dataset:han2024wildguard}, each paired with a specialized judging system and harm taxonomy. GPT-oss-120B serves as the general-purpose judge for red-team and simple-safety datasets; the LLaMA Guard 1–4 judges provide structured moderation through their respective taxonomies~\cite{llamaguard1:inan2312llama:2023,llamaguard2:2024,llama3:dubey2024llama:arxiv:2024,llamaguard4:2025}; and external moderation systems such as OpenAI Moderation and Perspective API evaluate toxicity, hate, and violence at scale~\cite{openai2024moderation,PerspectiveAPI}.

\noindent\textbf{Over Refusal.} This service evaluates whether a model can distinguish genuinely unsafe queries from benign ones, maintaining an appropriate balance between caution and responsiveness. We employ four tests which together expose excessive conservatism in safety-aligned models. \textit{OKTest}~\cite{dataset:oktest} presents clearly harmless prompts to measure baseline refusal accuracy. \textit{OR-Bench}~\cite{dataset:orbench} introduces near-safe queries across ten semantic categories (e.g., deception, harassment, hate, self-harm, sexual, violence) to assess whether safety filters overgeneralize across sensitive topics. \textit{XSTest}~\cite{dataset:xstest} probes exaggerated refusals arising from linguistic ambiguity, using paired categories such as homonyms, figurative language, and historical events to test whether models differentiate safe from unsafe meanings within contextual nuance. \textit{WildGuard}~\cite{dataset:han2024wildguard} examines real-world prompt distributions where models frequently decline legitimate queries due to overactive safety mechanisms.

\subsection{Security \& Robustness}

The middle layer assesses whether a model's security and reliability persist under intentional or accidental stressors.

We operationalize this through three services: \textit{Adversarial Robustness}, which measures resistance to diverse prompt-level attacks; \textit{Jailbreak Robustness}, which evaluates the model's ability to enforce alignment boundaries under adversarial instructions; and \textit{Out-of-Distribution Robustness}, which tests generalization beyond the training distribution.

\noindent\textbf{Adversarial Robustness.} This service evaluates how effectively a model resists prompt perturbations designed to induce incorrect, unsafe, or inconsistent responses. We employ two tests, \textit{AdvGLUE}~\cite{dataset:advglue} and \textit{AdvGLUE++}~\cite{decodingtrust:wang2023decodingtrust:neurips:2023}, which extend the GLUE~\cite{wang2019gluemultitaskbenchmarkanalysis} test suite with adversarially modified natural language inference, paraphrasing, and sentiment analysis tasks. These tests probe model brittleness through subtle lexical substitutions, negations, and distractor phrases that preserve grammatical validity while altering semantic intent.

\noindent\textbf{Jailbreak Robustness.} This service assesses whether a model upholds its alignment boundaries when confronted with direct, obfuscated, or adaptive attempts to elicit unsafe behavior. We integrate three complementary tests capturing distinct attack modes. \textit{Jailbroken}~\cite{wei2023jailbroken}, based on the AdvBench dataset~\cite{dataset:advbench:zou2023:arxiv:2023}, targets prompt-level exploits that induce safety violations through manipulative instructions or roleplay. \textit{Cipher}~\cite{yuan2024cipher} encodes unsafe intent using alternative linguistic or symbolic representations (e.g., Base64 or substitution ciphers), evaluating whether safety filters extend beyond surface-text recognition. \textit{Pair}~\cite{chao2024pair} employs an attacker LLM that dynamically rewrites jailbreak prompts, probing whether defenses remain consistent against adaptive, context-sensitive adversaries.

\noindent\textbf{Out-of-Distribution Robustness.} This service evaluates how well a model maintains consistent behavior when inputs deviate from its familiar linguistic or stylistic distributions. We employ the \textit{DecodingTrust}~\cite{decodingtrust:wang2023decodingtrust:neurips:2023} test, which transforms sentiment-analysis texts into diverse styles (Shakespearean, biblical, romantic, tweet) and applies common augmentations (e.g., typos, whitespace, etc.) while preserving their underlying semantics. Each variant is further perturbed at multiple intensity levels to simulate realistic shifts in expression, vocabulary, and tone.

\subsection{Privacy \& Fairness}

The top layer evaluates whether a model behaves equitably and protects sensitive information across diverse users and data contexts. It focuses on societal reliability, ensuring that outputs remain unbiased, inclusive, and privacy-preserving across demographic, ideological, and contextual variations.
We achieve this through two services: \textit{Model \& Data Privacy}, which measures the exposure and memorization of sensitive information, and \textit{Fairness \& Bias}, which evaluates consistency and impartiality across social groups.

\noindent\textbf{Model \& Data Privacy.} This service measures how effectively a model safeguards sensitive information and adheres to privacy expectations across both data-level and behavioral dimensions. We employ four complementary tests: \textit{Enron}, to examine memorization-based leakage by prompting models to reproduce email addresses~\cite{enron:klimt2004enron:2004}; ECHR to extend the analysis to legal text continuations, testing whether models regenerate personally identifiable information (PII) such as names, dates, or locations~\cite{echr:poudyal2020echr}; \textit{PII Awareness}~\cite{trustllm:huang2024:2024:arxiv} and \textit{ConfAIDE}~\cite{confaide:mireshghallah2023can:arxiv:2023} to evaluate reasoning-based privacy behavior. \textit{PII Awareness} tests whether models recognize privacy-sensitive queries and appropriately refuse to disclose items such as bank details or home addresses, both with and without explicit policy reminders. \textit{ConfAIDE} compares model judgments on privacy scenarios against human-labeled norms, quantifying alignment with societal privacy values.

\noindent\textbf{Fairness \& Bias.} This service evaluates whether a model produces impartial outputs across social, demographic, and ideological dimensions. We employ four tests spanning reasoning, classification, and value alignment. \textit{BBQ}~\cite{dataset:bbq:parrish2021:arxiv:2021} measures stereotype bias in contextual reasoning through paired demographic scenarios requiring unbiased answers. \textit{Disparagement}, derived from the UCI Adult dataset~\cite{dataset:adult}, examines fairness in income and occupation predictions across protected attributes (e.g., gender, race, age), using association metrics such as Cramer's V. \textit{GenderCARE}~\cite{gendercare:tang2024gendercare:ccs:2024} probes gender bias in responses, testing whether models overuse gendered terms or pronouns in neutral contexts. \textit{Preference}~\cite{trustllm:huang2024:2024:arxiv} assesses normative bias by analyzing whether models favor specific ideologies or moral viewpoints.

\section{Experiments}
\label{sec:experiments}
\subsection{Experimental Setup}

We evaluated 121 LLMs spanning both proprietary and open-weight ecosystems, executing over 5,000 black-box examinations across the 46 tests and 9 services defined in our \framework framework. To ensure fair and reproducible comparisons, all evaluations were performed under identical inference conditions: API access for proprietary systems and containerized local deployments for open-weight models. Appendix~\ref{appendix:setup} details the inference and judge configurations, infrastructure specifications, and proprietary model versions used throughout the evaluation. All results, including detailed per-model reports, are publicly available through the publicly-accessible \system portal.

\textbf{Model Selection.}
To systematically address our research questions, our model selection covers a representative spectrum of architectures, scales, and alignment paradigms, chosen based on their rankings on public leaderboards~\cite{lmsysleaderboard:chiang2024chatbot:arxiv:2024}:
\begin{itemize}[leftmargin=*]
\itemsep0em
\item \textbf{Proprietary Systems.} We include frontier models such as OpenAI’s GPT-4o/5, Anthropic’s Claude-3.5/4.5, xAI’s Grok-3/4, Google’s Gemini-2.5/3, and DeepSeek-R1.
\item \textbf{Open-Weight Families.} The open-weight group encompasses leading models from OpenAI (GPT-oss), Meta (Llama-3.x/4), Alibaba (Qwen3.5), Google (Gemma3/4), and Microsoft (Phi-4). Our selection includes models with diverse training techniques and architectures, including distillation-based derivatives (e.g., from DeepSeek-R1 and Cogito-v1). Model sizes range from compact 0.8B parameter models to large-scale systems with up to 120B parameters.
\end{itemize}

\subsection{Cross-Service Leaderboard and Initial Insights}

Table~\ref{tab:leaderboard} presents the top~20 models ranked by overall score. GPT-5 (81.8), Gemini~3 (80.6), Gemma4-31B (80.0), and Claude-4.5 (80.0) lead the ranking, but no single model dominates all dimensions. Gemma4-31B achieves the highest Fairness \& Bias (79.0), GPT-4o the highest Privacy (91.1), Qwen3.5-27B the highest Jailbreak Robustness (99.5), and Claude-3.5 the highest Safety Alignment (99.3). Notably, open-weight models rank alongside proprietary ones: Gemma4-31B ties Claude-4.5 for third place, and GPT-120B (78.7) outperforms GPT-4o (78.0). The spread between the top-ranked and 20th-ranked model is 8.4 points, yet the service-level variance within individual models is far larger, underscoring that aggregate scores mask critical dimensional trade-offs. 
% Extended results including distillation analyses are in Appendix~\ref{app:full_leaderboard}.

\begin{table}[t]
    \centering
    \caption{\framework evaluation scores for the top 20 models across nine services. No model dominates all dimensions: models exhibit distinct strength-weakness profiles, with trade-offs between safety, robustness, privacy, and utility. Extended results are in Appendix~\ref{app:full_leaderboard}; the complete leaderboard is available through the \system portal.}
    \resizebox{\linewidth}{!}{
    \begin{tabular}{l@{\hskip 4pt} @{\hskip 4pt}c@{\hskip 4pt}c@{\hskip 4pt}c@{\hskip 4pt}c@{\hskip 4pt}c@{\hskip 4pt}c@{\hskip 4pt}c@{\hskip 4pt}c@{\hskip 4pt}c@{\hskip 4pt} @{\hskip 4pt}c}
        \toprule
        & \multicolumn{8}{c}{\textbf{Services}} & \\
        \cmidrule(lr){2-10}
         \textbf{Model}
        & \shortstack{\textbf{Adv.}\\\textbf{Robust.}}
        & \shortstack{\textbf{Code}\\\textbf{Sec.}}
        & \shortstack{\textbf{Fair.}\\\textbf{\& Bias}}
        & \shortstack{\textbf{Halluc.}}
        & \shortstack{\textbf{Jail.}\\\textbf{Robust.}}
        & \shortstack{\textbf{M\&D}\\\textbf{Priv.}}
        & \shortstack{\textbf{OOD}\\\textbf{Robust.}}
        & \shortstack{\textbf{Over}\\\textbf{Ref.}}
        & \shortstack{\textbf{Safety}\\\textbf{Align.}}
        & \shortstack{\textbf{Overall} \\ \textbf{Score}} \\
        \midrule
\includegraphics[width=0.33cm]{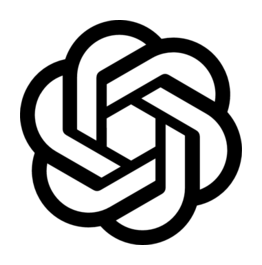} GPT-5 & 71.6 & 79.7 & 68.8 & \textbf{83.9} & 90.2 & 69.4 & 89.5 & 87.9 & 95.6& \textbf{81.8} \\
\includegraphics[width=0.33cm]{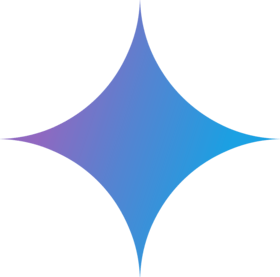} Gemini 3 & \textbf{77.2} & 74.3 & 76.7 & 82.7 & 70.3 & 65.1 & \textbf{91.6} & \textbf{94.4} & 93.2& \textbf{80.6} \\
\includegraphics[width=0.33cm]{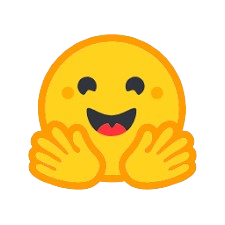} Gemma4-31B & 75.2 & 75.3 & \textbf{79.0} & 75.7 & 73.6 & 69.6 & 89.1 & 85.3 & 97.4& \textbf{80.0} \\
\includegraphics[width=0.33cm]{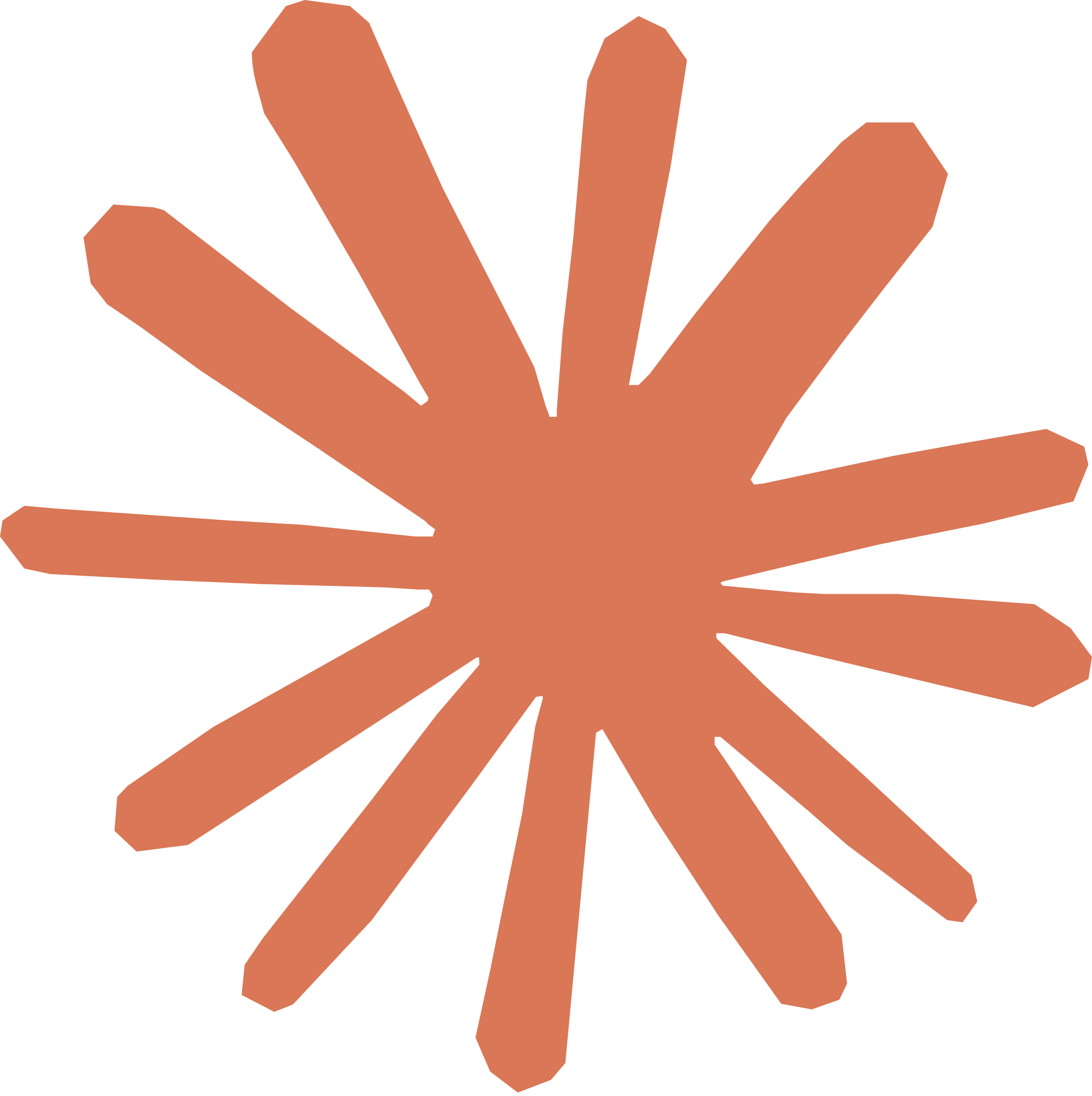} Claude-4.5 & 61.6 & 70.3 & 62.7 & 80.8 & 98.5 & 78.7 & 88.1 & 82.3 & 97.0& \textbf{80.0} \\
\includegraphics[width=0.33cm]{images/logo/huggingface-logo.png} Gemma4-26B-A4B & 70.1 & 76.4 & 77.5 & 73.2 & 72.7 & 70.8 & 89.5 & 82.3 & 97.3& \textbf{78.9} \\
\includegraphics[width=0.33cm]{images/logo/huggingface-logo.png} GPT-120B & 66.3 & \textbf{80.1} & 72.9 & 64.2 & 91.9 & 76.9 & 85.0 & 72.4 & 98.2& \textbf{78.7} \\
\includegraphics[width=0.33cm]{images/logo/openai-logo.png} GPT-4o & 62.6 & 73.1 & 67.5 & 77.5 & 51.4 & \textbf{91.1} & 86.9 & \textbf{94.4} & 97.5& \textbf{78.0} \\
\includegraphics[width=0.33cm]{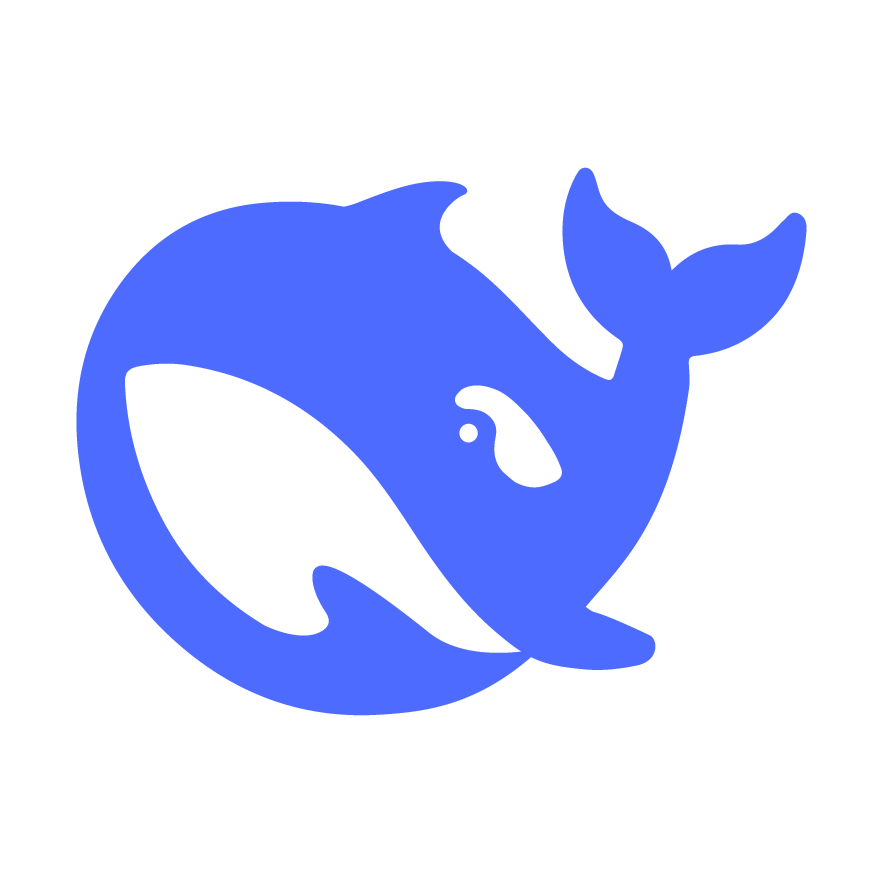} DeepSeek-R1 & 67.2 & 77.1 & 61.6 & 74.9 & 51.8 & 84.7 & 86.2 & 91.5 & 97.4& \textbf{76.9} \\
\includegraphics[width=0.33cm]{images/logo/huggingface-logo.png} Qwen3.5-27B & 69.8 & 75.0 & 71.2 & 71.5 & \textbf{99.5} & 48.4 & 86.5 & 79.6 & 85.6& \textbf{76.3} \\
\includegraphics[width=0.33cm]{images/logo/huggingface-logo.png} GPT-20B & 65.9 & 78.2 & 67.6 & 59.7 & 87.7 & 72.0 & 83.7 & 74.5 & 97.5& \textbf{76.3} \\
\includegraphics[width=0.33cm]{images/logo/huggingface-logo.png} Qwen3.5-122B-A10B & 69.2 & 74.1 & 64.3 & 73.1 & 99.2 & 54.1 & 86.9 & 79.0 & 85.4& \textbf{76.1} \\
\includegraphics[width=0.33cm]{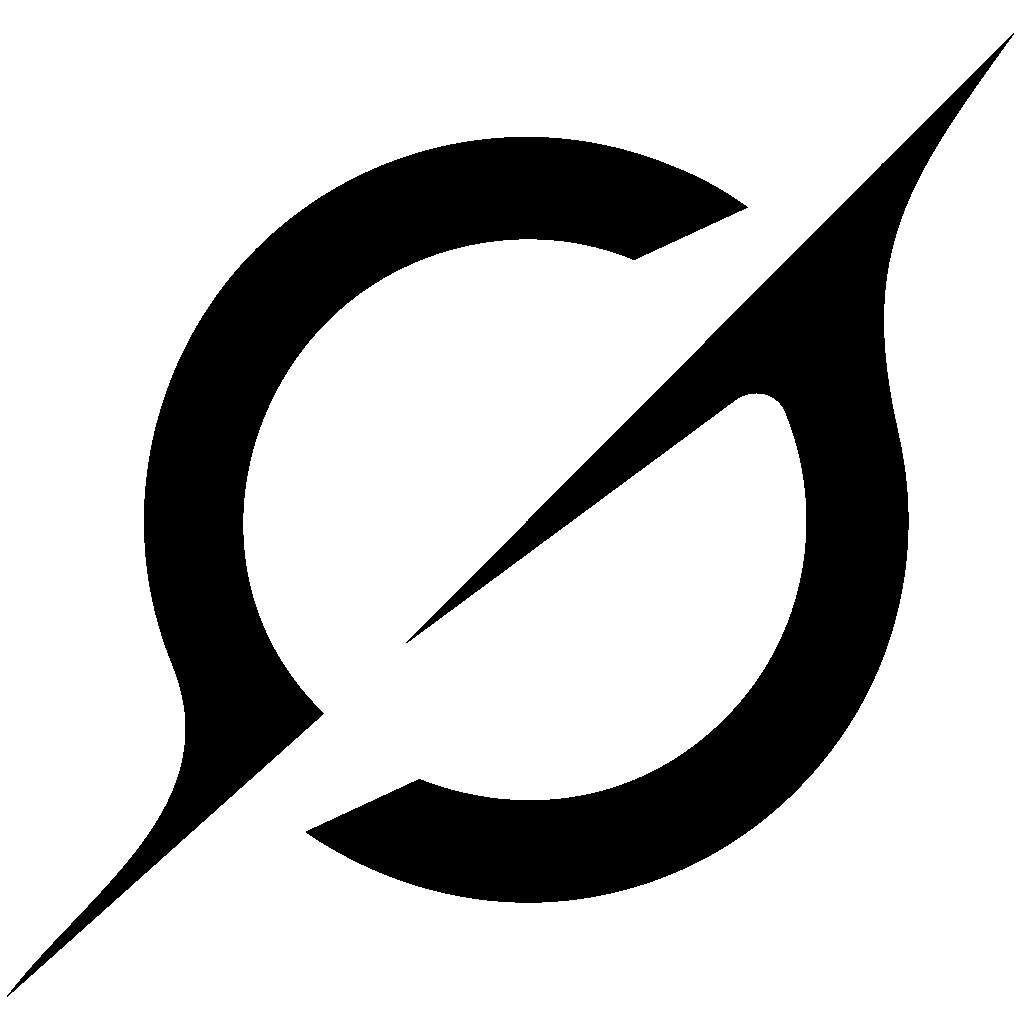} Grok 4 & 70.3 & 69.7 & 61.5 & 66.0 & 73.2 & 68.4 & 90.0 & 88.3 & 97.0& \textbf{76.0} \\
\includegraphics[width=0.33cm]{images/logo/grok-logo-light.png} Grok 4.1 & 68.3 & 72.0 & 59.5 & 66.1 & 78.1 & 68.2 & 89.8 & 88.2 & 88.0& \textbf{75.3} \\
\includegraphics[width=0.33cm]{images/logo/claude-logo-light.png} Claude-3.5 & 62.7 & 72.2 & 61.5 & 74.9 & 70.5 & 80.8 & 87.0 & 67.2 & \textbf{99.3}& \textbf{75.1} \\
\includegraphics[width=0.33cm]{images/logo/huggingface-logo.png} Phi-4 & 63.5 & 68.0 & 72.0 & 58.4 & 56.8 & 84.2 & 87.3 & 86.3 & 99.0& \textbf{75.1} \\
\includegraphics[width=0.33cm]{images/logo/huggingface-logo.png} Qwen3.5-35B-A3B & 68.1 & 72.1 & 63.0 & 70.6 & 98.7 & 54.0 & 84.9 & 78.4 & 85.5& \textbf{75.0} \\
\includegraphics[width=0.33cm]{images/logo/huggingface-logo.png} Qwen3.5-9B & 66.1 & 73.2 & 63.8 & 70.5 & 98.8 & 52.7 & 84.1 & 78.7 & 86.6& \textbf{74.9} \\
\includegraphics[width=0.33cm]{images/logo/huggingface-logo.png} Llama3.3-70B & 67.6 & 68.4 & 58.9 & 72.1 & 47.8 & 74.4 & 88.6 & 94.0 & 94.1& \textbf{74.0} \\
\includegraphics[width=0.33cm]{images/logo/grok-logo-light.png} Grok 3 & 66.3 & 68.6 & 60.9 & 74.0 & 47.4 & 67.4 & 89.6 & \textbf{94.4} & 94.0& \textbf{73.6} \\
\includegraphics[width=0.33cm]{images/logo/huggingface-logo.png} Qwen3.5-4B & 65.1 & 71.5 & 59.3 & 68.8 & 98.6 & 51.0 & 82.4 & 78.5 & 85.2& \textbf{73.4} \\
        \bottomrule
    \end{tabular}
    }
    \label{tab:leaderboard}
\end{table}

\subsection{RQ1: What is the Trade-Off Between Safety and Utility?}

To move beyond aggregate scores, we first quantify the systemic relationships between our nine evaluation services and then visualize how different models navigate the resulting trade-offs.

\begin{figure}[t]
    \centering
    \includegraphics[width=0.6\columnwidth]{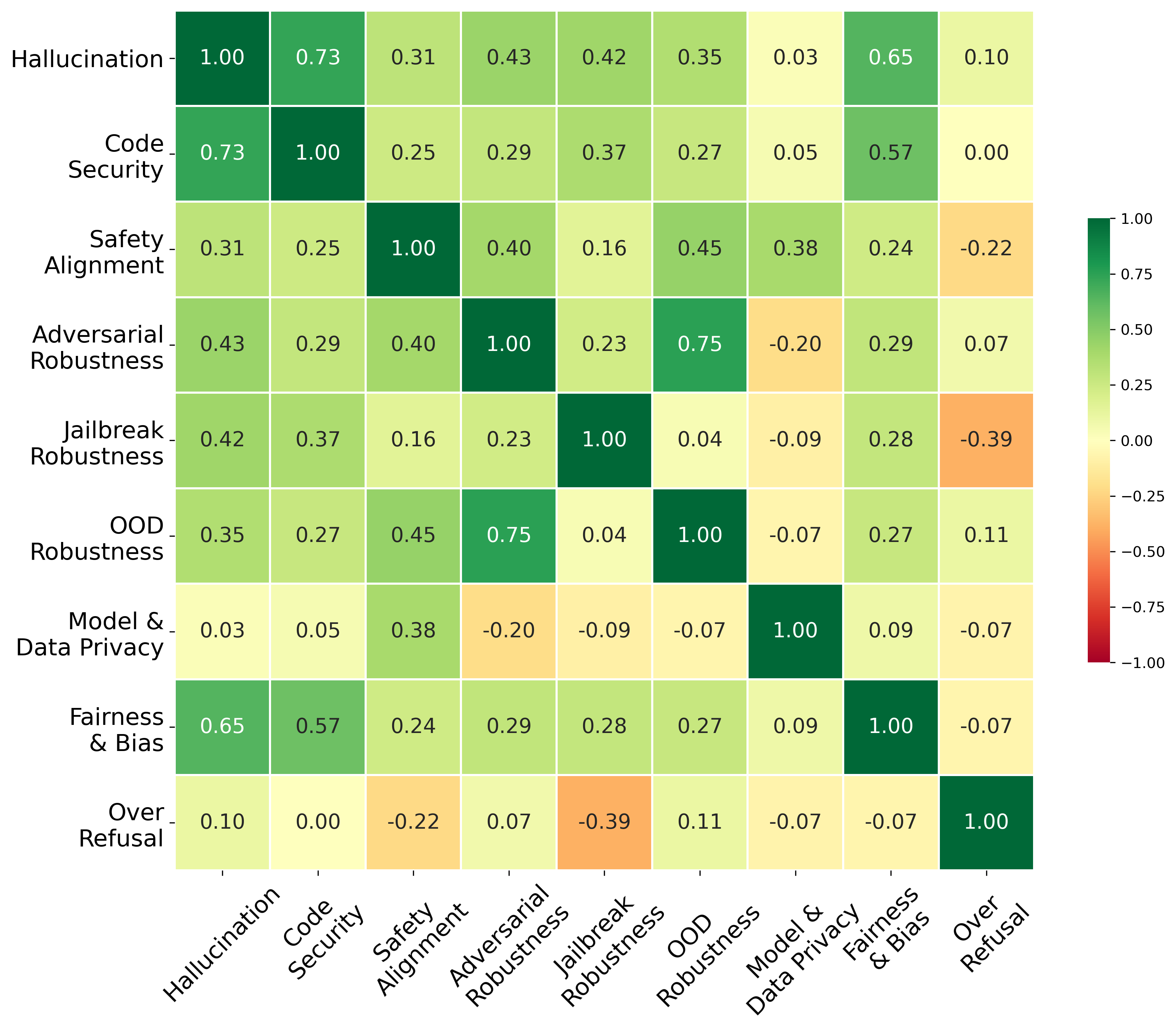}
\caption{Cross-service correlation and dependency analysis.}
    \label{fig:correlation-matrix}
\end{figure}

\noindent\textbf{Safety enforcement trades off against utility, while privacy is largely orthogonal to other dimensions.}
We computed the correlation matrix across the full dataset of over 5,000 examination results (Figure~\ref{fig:correlation-matrix}). The analysis reveals three distinct patterns. First, services within the Safety \& Reliability cluster (e.g., Hallucination and Code Security, $r{=}0.73$, 95\% CI: [0.62, 0.81]) and the Security \& Robustness cluster (e.g., Adversarial and OOD Robustness, $r{=}0.75$, 95\% CI: [0.65, 0.83]) are strongly self-correlated, suggesting that improvements in core capabilities lead to simultaneous gains across related areas. Second, the most significant conflict is the negative correlation between Over Refusal and safety-centric services, strongest for Jailbreak Robustness ($r{=}{-}0.39$, 95\% CI: [$-$0.54, $-$0.20]) and Safety Alignment ($r{=}{-}0.22$, 95\% CI: [$-$0.40, $-$0.02]), empirically quantifying the \textit{safety tax}: as models improve at refusing harmful content, they get worse at correctly answering benign prompts. Third, Model \& Data Privacy exhibits weak correlations with most other services (mean $|r|{=}0.13$, 95\% CI: [0.06, 0.21]; six of eight pairwise correlations are non-significant), with a moderate positive exception for Safety Alignment ($r{=}0.38$, 95\% CI: [0.20, 0.54]). This provides strong evidence that privacy is largely orthogonal to other trustworthiness dimensions and is not addressed by standard safety alignment.

\begin{figure}
    \centering
    \includegraphics[width=\linewidth]{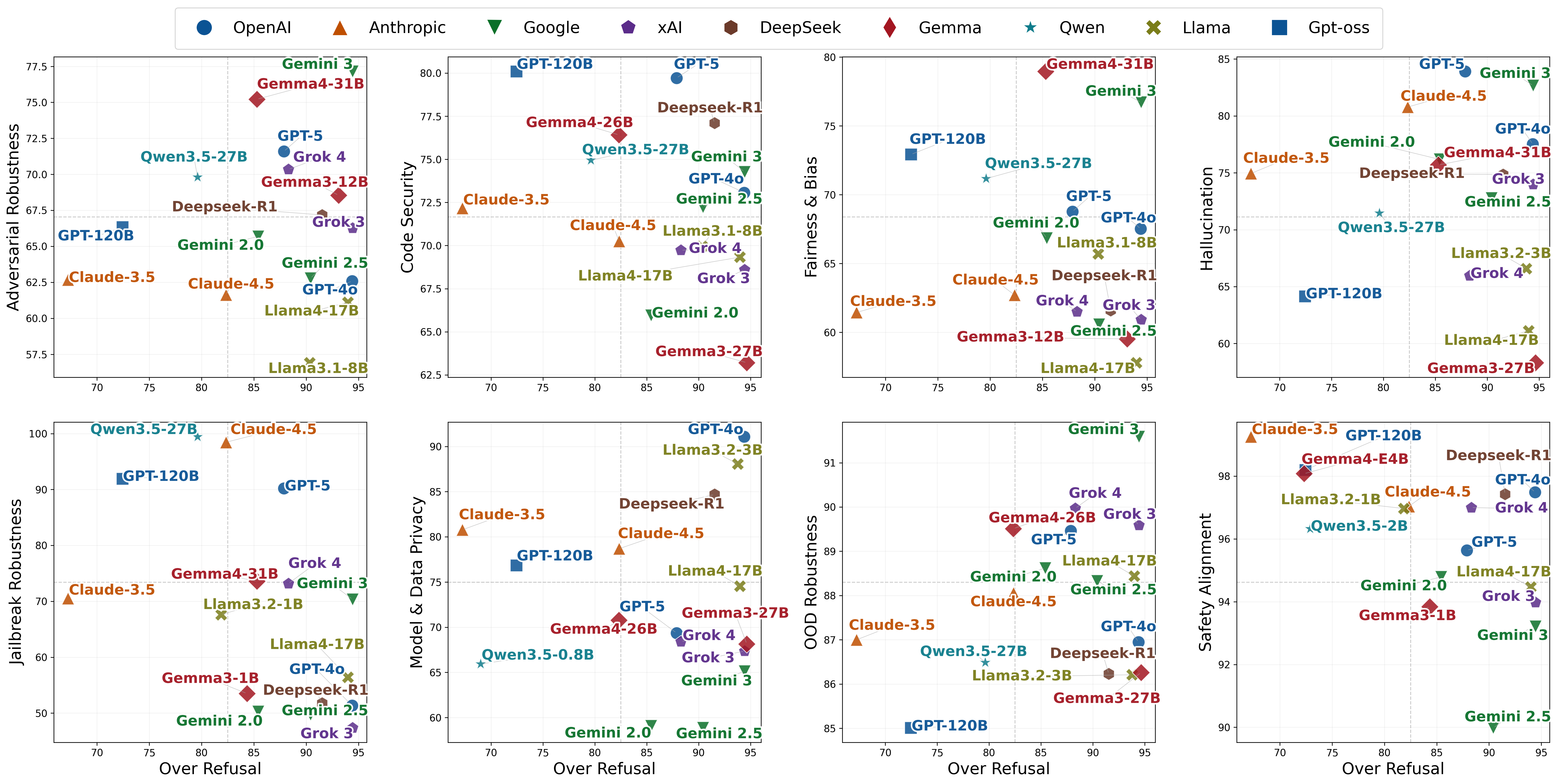}
    \caption{Safety vs. Utility Tension Analysis. Model performance for eight key services ($y$-axis) against the Over Refusal ($x$-axis) score. A higher Over Refusal score means the model refuses fewer benign prompts. The dashed lines indicate the mean performance on each axis, dividing the models into four quadrants. The top-right quadrant represents the ideal state of high performance and high utility. The plots visually demonstrate the \textit{safety tax}, where models with the highest safety (e.g., Claude-4.5, GPT-120B) have the lowest utility. Conversely, services like Model \& Data Privacy show that high performance and high utility can be achieved simultaneously.}
    \label{fig:over_refusal_scatterbox}
\end{figure}

Figure~\ref{fig:over_refusal_scatterbox} visualizes how individual models resolve this tension. The Jailbreak Robustness plot is representative: the desirable top-right quadrant (high robustness, high utility) is nearly empty. Instead, models cluster into two camps. The ``Safety-First'' camp (Claude-4.5, GPT-120B) achieves strong Jailbreak Robustness and Fairness but records the lowest Over Refusal scores. The ``Utility-First'' camp (Gemini~3, Grok~4, Llama4-17B-16E) maximizes utility at the cost of significantly weaker jailbreak and adversarial resistance.

The \raid{near-orthogonality of privacy (mean $|r|{=}0.13$)} is particularly striking. Qwen3.5-27B leads the ecosystem in Jailbreak Robustness (99.5) yet records the lowest privacy among open-weight models (48.4), while GPT-4o achieves the highest privacy of any model (91.1) despite weak jailbreak resistance (51.4). When we turn to the Model \& Data Privacy plot, the trade-off disappears: the desirable top-right quadrant is now populated, with models like Llama3.2-3B and DeepSeek-R1 demonstrating both high privacy and high utility. This distinction suggests that the \textit{safety tax} is a consequence of blunt, refusal-based alignment strategies, whereas privacy is achievable through methods that do not damage utility, such as improved data curation and PII removal during pre-training~\cite{decodingtrust:wang2023decodingtrust:neurips:2023}.

\noindent\textbf{Deployment context reshapes rankings.}
The trade-offs documented above are not merely theoretical: they materially alter which model is ``best'' depending on deployment priorities. To quantify this, we define six \emph{stakeholder profiles}, normalized weight vectors over the nine services that capture the priorities of representative deployment contexts (e.g., Healthcare emphasizes hallucination and privacy; Code~Assistant emphasizes code security and adversarial robustness). Among the top~15 models, the identity of the top model changes across four of six profiles, mean pairwise Kendall's~$\tau$ drops to~0.455 (95\% CI: [0.33, 0.57]), and 27\% of pairwise model orderings reverse when switching perspectives. The most sensitive model, GPT-4o, swings 13 rank positions between its best and worst profiles. A Dirichlet robustness analysis over 10{,}000 random weight vectors confirms this is not an artifact of the chosen profiles: while coarse tier membership is stable ($\tau{=}0.775$ vs.\ Uniform), within-tier ordering remains sensitive to the evaluation perspective. These results confirm that a single aggregate ranking cannot capture deployment-specific trustworthiness; full profile definitions and per-model rankings are in Appendix~\ref{sec:stakeholder} (Table~\ref{tab:stakeholder_scores_ranks}).

\begin{keyfinding}
\raid{\textbf{Summary.} Safety and utility remain in tension ($r{=}{-}0.39$), while privacy is largely independent of other dimensions (mean $|r|{=}0.13$). Deployment priorities can shift model rankings by up to 13 positions, showing that no single aggregate score captures deployable trustworthiness.}
\end{keyfinding}

\subsection{RQ2: What is the Impact of Scale and Alignment on Safety?}

To disentangle scale from alignment, we perform two controlled comparisons: (i)~intra-family comparisons within Qwen3/3.5, Gemma3, and Llama3, where models share the same architecture and post-training~\cite{team2025gemma3:technicalreport,yang2025qwen3:technicalreport,llama3:dubey2024llama:arxiv:2024} and differ only in size; and (ii)~cross-distillation comparisons of models sharing the same base architecture but fine-tuned with different pipelines (DeepSeek-R1, Cogito/IDA).

\begin{figure}
    \centering
    \includegraphics[width=0.6\columnwidth]{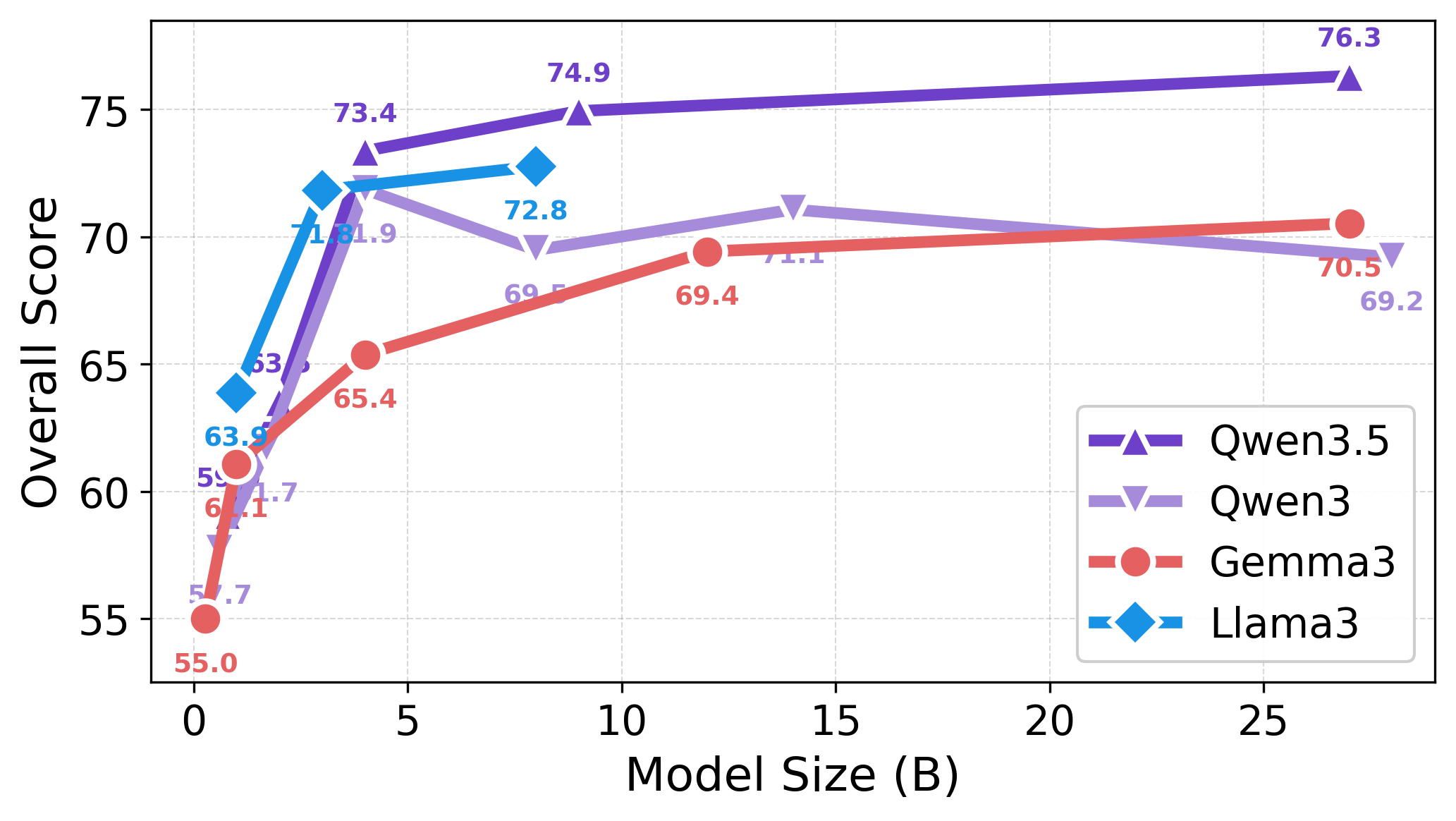}
\caption{The performance impact of model size for open-weight families (Qwen, Gemma, Llama) that share the same post-training within each family. The plot shows a general upward trend with scale but highlights diminishing returns beyond $\sim$4B parameters.}
    \label{fig:model_size_comparison_line}
\end{figure}

\noindent\textbf{Alignment strategy, not scale alone, drives a model's safety profile.}
Our analysis reveals that the popular assumption of ``bigger is better" for safety is an oversimplification. Figure~\ref{fig:model_size_comparison_line} plots the overall scores for the Qwen3/3.5, Gemma3, and Llama3 families against their size. While there is a general upward trend, the relationship is clearly not linear, revealing a consistent two-phase pattern. Initially, performance increases almost linearly with size up to the $\sim$4B parameter mark. For instance, Qwen3.5 jumps from 59.0 (0.8B) to 73.4 (4B), a gain of over 14 points, but then plateaus with only 3 additional points from 4B to 27B. We observe a consistent plateau around the 4B scale. Beyond this point,
models enter a phase of diminishing returns. This consistent trend across the model families examined suggests that a minimum model scale is a necessary prerequisite for learning complex safety behaviors. Below this threshold, models may lack the capacity to fully generalize safety principles. Above it, however, simple parameter growth is insufficient.

\noindent\textbf{The dominant role of alignment strategy.}
Our findings show that the post-training alignment strategy is the primary determinant of a model's position on the safety-utility spectrum. A clear example is the behavior of GPT-120B. While built on an open-weight architecture, its aggressive safety tuning places it far from its open-weight peers (like Llama and Qwen) and squarely in the ``Safety-First" camp alongside proprietary models like Claude-4.5. This demonstrates that a model's safety posture is less a function of its base capabilities and more a direct outcome of the priorities encoded during its alignment phase.

\begin{figure}
    \centering
    \includegraphics[width=0.8\columnwidth]{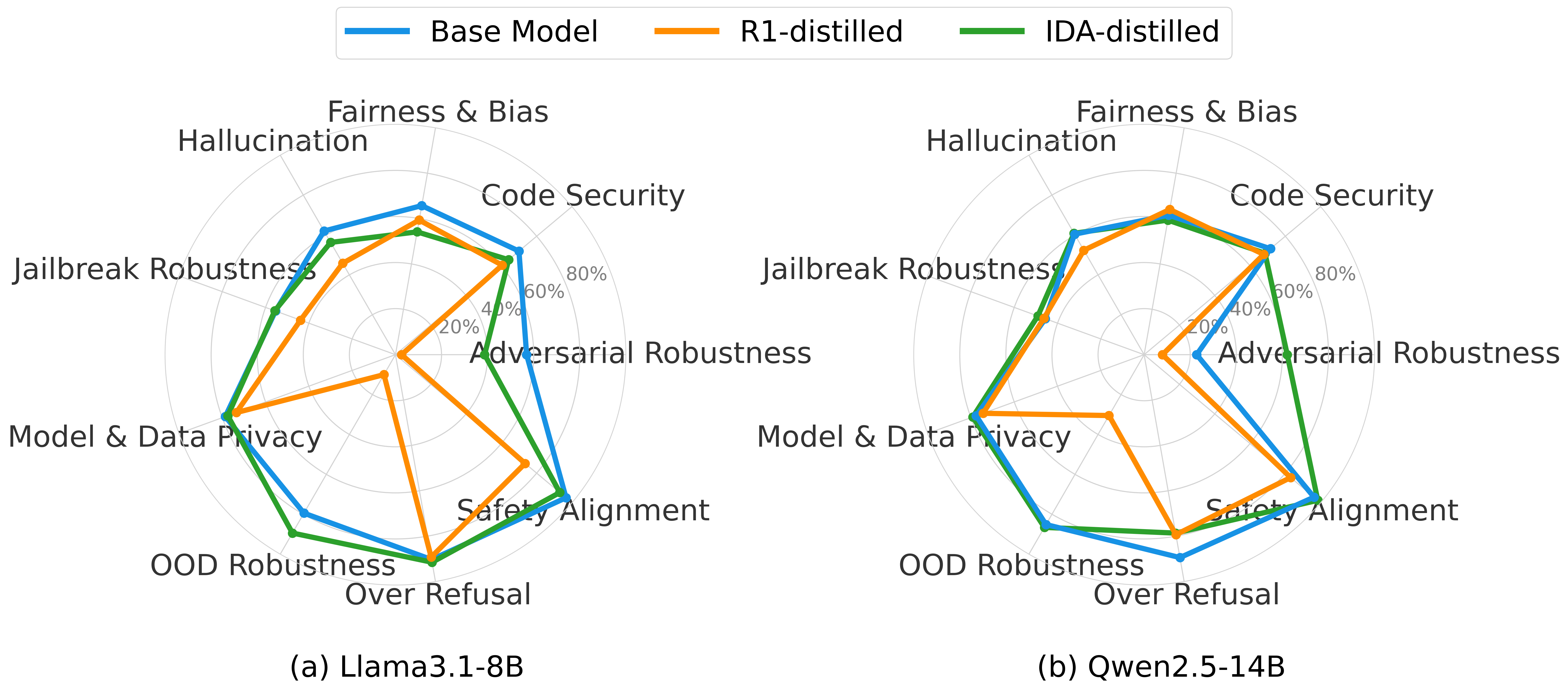}
\caption{The impact of different distillation methods on (a) Llama3.1-8B and (b) Qwen2.5-14B base models. The radar plots highlight the dramatic and divergent effects of R1 distillation in Adversarial and OOD Robustness, while the IDA method has a more nuanced, and in the case of Qwen, positive impact.}
    \label{fig:distilation_comparison_radar}
\end{figure}

To isolate this effect further, we compare models sharing the same base architecture but differing in alignment strategy. Safety and utility profiles diverge dramatically, often more so than between models from entirely different ecosystems.
Figure~\ref{fig:distilation_comparison_radar} visualizes this effect, comparing base models (Llama3.1-8B and Qwen2.5-14B) against their distilled derivatives. Distillation methods that prioritize reasoning (e.g., DeepSeek-R1) or instruction-following (e.g., Cogito/IDA) introduce markedly different safety profiles. For Qwen2.5-14B, Cogito’s IDA distillation improves adversarial robustness relative to the base model, whereas R1 distillation causes a sharp collapse. The divergence is even more pronounced for Llama-3.1-8B. While IDA moderately reduces robustness, R1 distillation reduces adversarial robustness to near-zero, demonstrating that alignment strategies can radically alter robustness.
This demonstrates that while knowledge can be distilled, the process must preserve the nuanced decision boundaries required for robustness, a property that current pipelines do not guarantee.

We trace this collapse to a structural issue in off-policy distillation. In R1-style models, the on-policy KL correction~\cite{agarwal2024policy} is omitted due to vocabulary mismatches between teacher and student, causing the student to overfit exact teacher trajectories. This collapses the entropy of the predictive distribution, producing sharp decision boundaries that are brittle to small input perturbations~\cite{goodfellow2014explaining}. Appendix~\ref{app:distillation-embeddings} formalizes this mechanism, contrasting the on-policy and off-policy objectives and showing how the omission of on-policy correction drives entropy collapse.

\begin{figure}[t]
    \centering
    \includegraphics[width=\columnwidth]{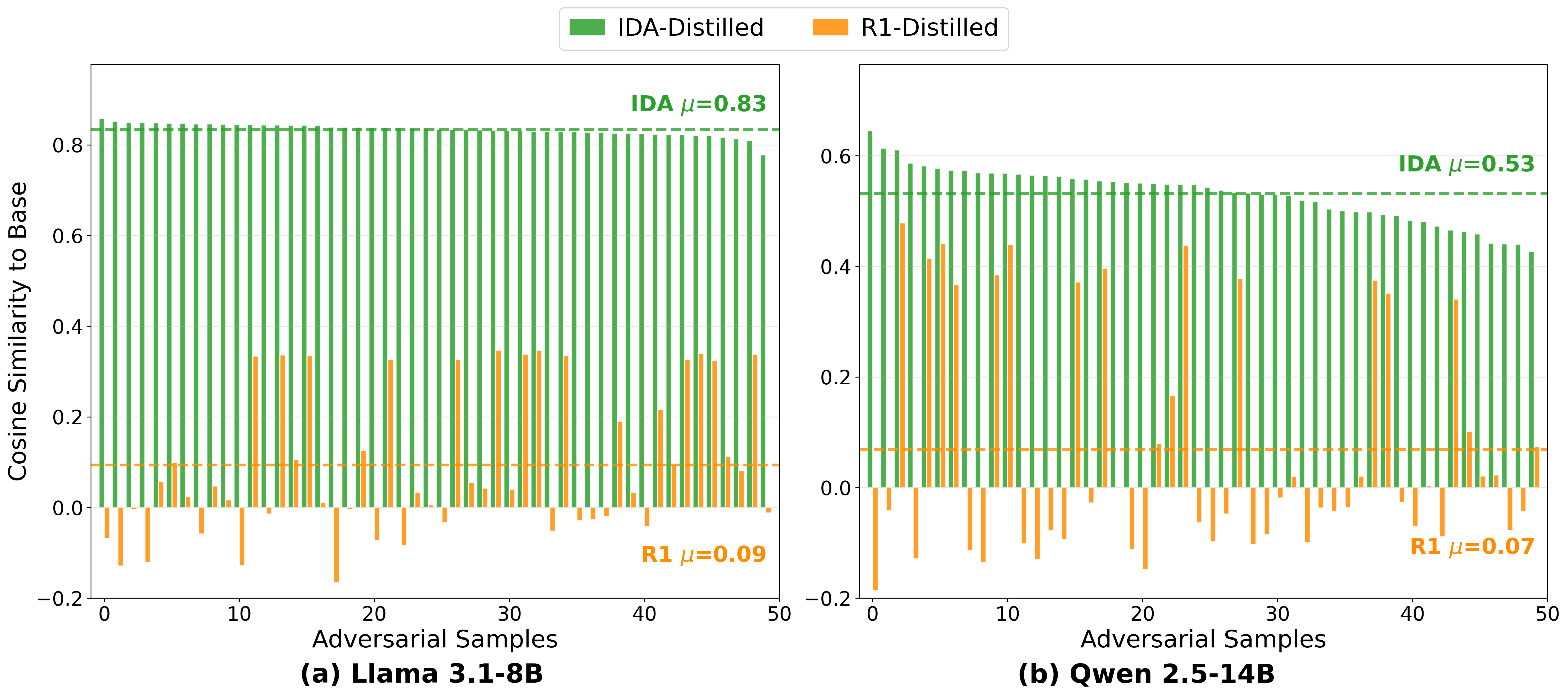}
    \caption{Per-sample cosine similarity between base and distilled model representations on 50 AdvGLUE-SST2 adversarial prompts. R1 distillation produces near-orthogonal representations, while IDA distillation maintains high alignment with the base model.}
    \label{fig:distillation-cosine}
\end{figure}

To corroborate this mechanism, we move beyond the black-box evaluation protocol and extract last-layer hidden-state representations from the open-weight base and distilled models on the same adversarial inputs. We sample 50 adversarial prompts from the AdvGLUE-SST2 corpus and extract the last-layer hidden state for each base and distilled model at the final input position, yielding per-sample embedding vectors. We then measure per-sample cosine similarity between each distilled model's embedding and the corresponding base-model embedding: high similarity indicates the distilled model preserves the base model's representational geometry; low similarity indicates divergence.

Figure~\ref{fig:distillation-cosine} shows that R1 distillation produces near-orthogonal representations relative to the base model ($\bar{\cos}{=}0.094{\pm}0.161$ for Llama, $0.069{\pm}0.206$ for Qwen), confirming that the off-policy objective fundamentally restructures the representation space. IDA retains substantially more structure ($\bar{\cos}{=}0.834{\pm}0.013$ for Llama, $0.532{\pm}0.048$ for Qwen; Cohen's $d{=}6.5$ and $3.1$ respectively), consistent with its on-policy correction preserving smoother decision boundaries. These results identify \emph{distillation-induced robustness collapse} as a structural vulnerability in current strong-to-weak distillation pipelines, the first robustness-focused characterization of this failure mode.

\begin{keyfinding}
\raid{\textbf{Summary.} Above the ${\sim}4$B-parameter threshold, alignment strategy --- not size --- drives a model's safety profile. On the same base architecture, R1 distillation reduces adversarial robustness from 56.9 to 2.6, revealing a severe robustness failure that capability scores alone would miss.}
\end{keyfinding}

\subsection{RQ3: How Does Model Capability Relate to Privacy?}

\noindent\textbf{Advancing model capability degrades privacy.}
While our analysis of scale within a single generation reveals a complex picture, \raid{examining the evolution of seven model families across two consecutive generations reveals a recurring pattern}. Table~\ref{tab:family-comparison} quantifies the changes in overall and privacy scores \raid{between consecutive versions of each family}.
The data reveals a recurring pattern across generations. The most dramatic example is the transition from GPT-4o to GPT-5, where a nearly 4-point gain in the overall \framework score is accompanied by a 21.73-point collapse in the privacy score. Similarly, \raid{Qwen ($-6.77$) and Llama ($-13.53$) show privacy regressions alongside overall improvements, while Gemma ($-4.10$) exhibits a joint decline in both overall and privacy scores}.

However, this trajectory is not universal. Grok~3$\to$4 improves both overall ($+2.62$) and privacy ($+1.04$), and Gemini~2.0$\to$2.5 shows negligible privacy change ($-0.21$) despite an overall shift. These exceptions suggest that privacy degradation is not an inherent cost of capability improvement but rather a consequence of specific training and data curation choices.

\begin{table}
    \centering
    \caption{Per-family comparison across model generations, showing changes in overall and privacy scores. Some comparisons span architectural changes (e.g., Llama3.2$\to$Llama4), yet privacy regression recurs in most families.}
    \begin{tabular}{l @{\hspace{0.5cm}} l @{\hspace{0.5cm}} l @{\hspace{0.5cm}} r @{\hspace{0.5cm}} r}
        \toprule
        \textbf{Family} & \textbf{Old Model} & \textbf{New Model} 
        & $\boldsymbol{\Delta}$ Overall & $\boldsymbol{\Delta}$ Privacy \\
        \midrule
        Claude & Claude-3.5 & Claude-4.5 & $+4.88$ & $-2.09$ \\
        ChatGPT & GPT-4o & GPT-5 & $+3.84$ & $-21.73$ \\
        Grok & Grok 3 & Grok 4 & $+2.62$ & $+1.04$ \\
        Gemini & Gemini 2.0 & Gemini 2.5 & $-0.80$ & $-0.21$ \\
        Qwen & Qwen2.5-14B & Qwen3-14B & $+3.26$ & $-6.77$ \\
        Llama & Llama3.2-3B & Llama4-17B-16E & $+1.20$ & $-13.53$ \\
        Gemma & Gemma2-27B & Gemma3-27B & $-2.78$ & $-4.10$ \\
        \bottomrule
    \end{tabular}
    \label{tab:family-comparison}
\end{table}

\begin{keyfinding}
\raid{\textbf{Summary.} Privacy often regresses across model generations: 5 of 7 transitions show declines, including a 21.73-point drop from GPT-4o to GPT-5 despite a 3.84-point overall gain. Grok's counterexample shows that this regression is not an inherent cost of capability improvement.}
\end{keyfinding}

\subsection{RQ4: How Does Category-Specific Safety Generalize?}

As service-level alignment improves safety, model performance should remain stable across the related categories within each test of the service. As defined in Section~\ref{subsec:eval_framework}, a category refers not to the high-level service but to its fine-grained categories within tests: different topics, programming languages, attack types, and demographic contexts.

\begin{figure}
    \centering
    \includegraphics[width=0.8\columnwidth]{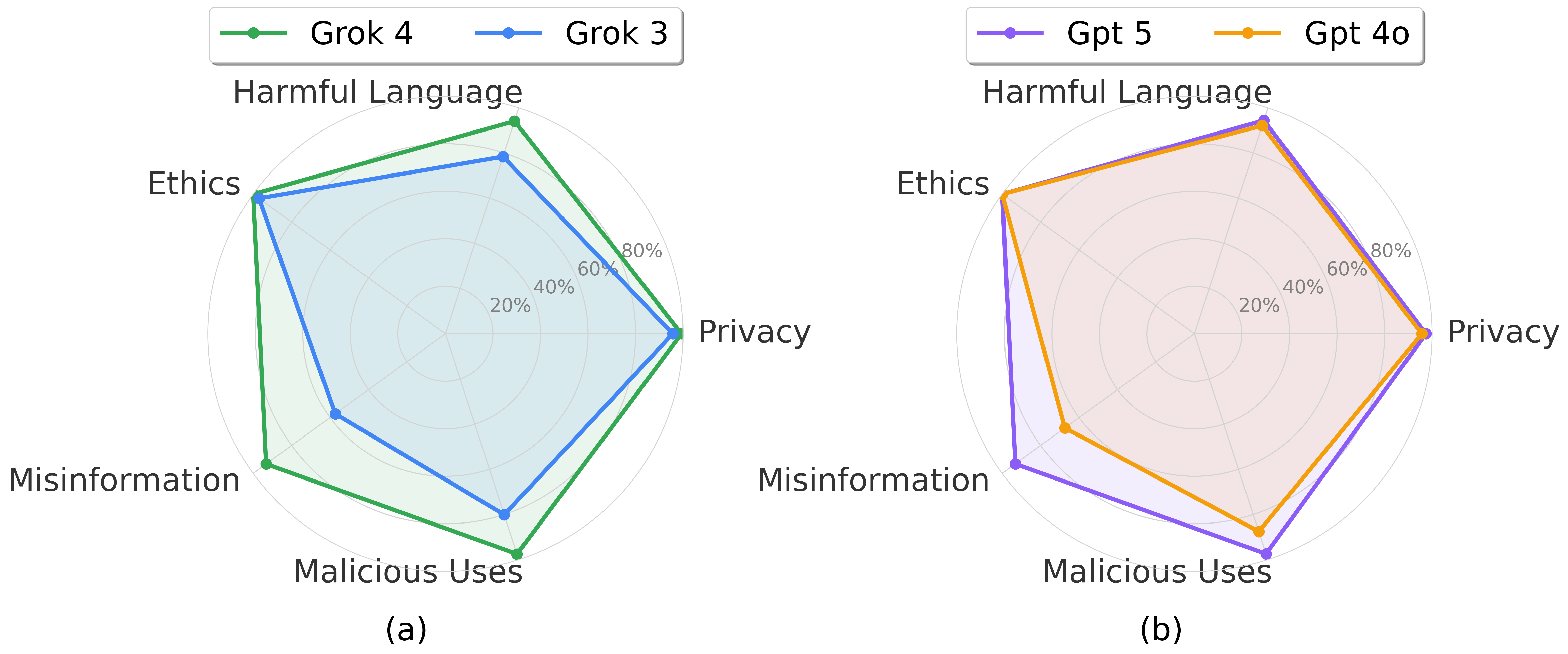}
    \caption{Generalization across five safety categories in the WildGuard test for (a) Grok and (b) GPT model families.}
    \label{fig:sample_categories_grok}
\end{figure}

However, as illustrated in Figure~\ref{fig:sample_categories_grok}, even frontier models display substantial variability. Grok~3, for instance, scores above 90 in two WildGuard safety categories but drops below 70 in three others, gaps that its successor partially closes. A similar pattern emerges in the GPT family, where the newer model reduces intra-test variance.

To quantify this, we analyze the intra-test performance variation across the 215 categories in our framework, with detailed results for each service provided in Appendix~\ref{app:service_results}. We measure two key metrics: the Average Standard Deviation of a model's scores across categories within a test, and the Average Generalization Gap, defined as the average difference between a model's best and worst category-level score. High values for either metric indicate non-generalizable behavior. Our analysis, aggregated to the service level in Table~\ref{tab:service-domain-generalization}, reveals that the performance is highly fragmented.

\noindent\textbf{Safety behaviors are category-dependent and fail to generalize, especially in security contexts.}
Our findings reveal a clear hierarchy of brittleness across the nine services. We find no evidence of a ``universally safe" model; instead, the degree of generalization is inversely related to the adversarial complexity of the service.
The services designed to test resilience against active attacks are by far the least generalizable. Jailbreak Robustness is the most fragmented service, with an average standard deviation of 17.8 points (95\% CI: [15.0, 18.2]) and a generalization gap of 47.0 points ([38.9, 47.1]): a model's ability to resist one style of jailbreak offers almost no guarantee of its ability to resist another. In contrast, OOD Robustness and Safety Alignment exhibit the lowest average standard deviations (4.4 [3.6, 4.6] and 5.6 [4.9, 6.0], respectively), with non-overlapping CIs confirming the hierarchy is statistically robust.

\noindent\textbf{Alignment as behavioral overfitting.}
These results suggest that current alignment methods function analogously to overfitting in supervised learning: they increase refusal rates on targeted safety scenarios but fail to generalize beyond the distributions represented in training data. Models remain vulnerable to adversarial paraphrasing, encoding transformations, and domain shifts, indicating that alignment enforces surface-level policy compliance rather than principle-driven risk understanding. This interpretation reframes over-refusal not merely as an optimization imbalance but as a symptom of distribution-specific learning, where models memorize familiar risk patterns without acquiring deeper contextual reasoning.

\begin{table}[t]
\centering
\caption{Service-level generalization statistics across test categories. Avg Std: mean standard deviation of a model's category-level scores within each test, averaged across models. Avg Gap: mean difference between a model's best and worst category score within each test. Higher values indicate more fragmented, category-dependent behavior.}
% \resizebox{\linewidth}{!}{
\begin{tabular}{l @{\hspace{0.5cm}} r @{\hspace{0.5cm}} r}
\toprule
\textbf{Service}  & \textbf{Avg Std (\%)} & \textbf{Avg Gap (\%)} \\
\midrule
Adversarial Robustness & 13.7 & 35.8 \\
Model \& Data Privacy  & 16.1 & 23.9 \\
Fairness \& Bias       & 7.2  & 18.4 \\
Code Security          & 9.4  & 17.3 \\
Hallucination          & 11.6 & 28.1 \\
Safety Alignment       & 5.6  & 16.8 \\
OOD Robustness         & 4.4  & 13.6 \\
Jailbreak Robustness  & 17.8 & 47.0 \\
Over Refusal           & 12.4 & 41.1 \\
\bottomrule
\end{tabular}
% }
\label{tab:service-domain-generalization}
\end{table}

\begin{keyfinding}
\raid{\textbf{Summary.} Safety generalization is highly fragmented: Jailbreak Robustness shows a 47-point within-service gap, versus 13.6 for OOD Robustness. This pattern is consistent with behavioral overfitting, where alignment learned for familiar risks fails to generalize to new ones.}
\end{keyfinding}

\section{Related Work}
\label{sec:relatedwork}

Recent work on LLM assurance has transitioned from capability benchmarks (e.g., MMLU~\cite{hendrycks2021measuringmassivemultitasklanguage}, BIG-Bench~\cite{luo2025bigbenchunifiedbenchmarkevaluating}) toward frameworks explicitly designed to measure safety, security, or trustworthiness. While these efforts differ in motivation and design, most assess isolated risk types, leaving significant fragmentation across evaluation domains.

To unify comparison, we organize prior works under the \framework taxonomy adopted by \system (Section~\ref{sec:framework}). This taxonomy separates behavioral integrity (how models act), adversarial resilience (how they fail), and societal reliability (how they treat people and data). Table~\ref{tab:coverage-by-service} summarizes how existing open-source frameworks and attack toolkits map to this taxonomy and the nine services composing \system.

\begin{table*}[t]
\centering
\caption{Coverage of major frameworks and toolkits across \system's nine services, grouped by the three evaluation dimensions. \priority{100} denotes explicit coverage; \priority{50} indicates partial or implicit coverage; \priority{0} means no coverage.}
\small
\setlength{\tabcolsep}{3pt}
\begin{adjustbox}{width=\textwidth}
\begin{tabular}{@{}l@{\hspace{0.25cm}}lcccccc c@{}}
\toprule
\textbf{Dimension} & \textbf{Service} &
\begin{tabular}{c}\textbf{Trust}\\\textbf{LLM}\end{tabular} &
\begin{tabular}{c}\textbf{Trust}\\\textbf{Eval}\end{tabular} &
\begin{tabular}{c}\textbf{Decoding}\\\textbf{Trust}\end{tabular} &
\begin{tabular}{c}\textbf{Prompt}\\\textbf{Bench}\end{tabular} &
\textbf{HELM} &
\textbf{garak} &
\begin{tabular}{c}\textbf{\system}\\\textbf{(ours)}\end{tabular} \\
\midrule
\multirow{4}{*}{\begin{tabular}[l]{@{}l@{}}Safety \&\\Reliability\end{tabular}}
  & Hallucination & \priority{0} & \priority{100} & \priority{50} & \priority{50} & \priority{0} & \priority{50} & \priority{100} \\
  & Code Security & \priority{0} & \priority{0} & \priority{0} & \priority{0} & \priority{0} & \priority{100} & \priority{100} \\
  & Safety Alignment & \priority{50} & \priority{100} & \priority{50} & \priority{100} & \priority{0} & \priority{100} & \priority{100} \\
  & Over Refusal & \priority{50} & \priority{50} & \priority{50} & \priority{0} & \priority{0} & \priority{0} & \priority{100} \\
\midrule
\multirow{3}{*}{\begin{tabular}[l]{@{}l@{}}Security \&\\Robustness\end{tabular}}
  & Adversarial Robustness & \priority{0} & \priority{100} & \priority{100} & \priority{100} & \priority{100} & \priority{100} & \priority{100} \\
  & Jailbreak Robustness & \priority{50} & \priority{50} & \priority{50} & \priority{50} & \priority{100} & \priority{100} & \priority{100} \\
  & OOD Robustness & \priority{0} & \priority{100} & \priority{50} & \priority{100} & \priority{0} & \priority{0} & \priority{100} \\
\midrule
\multirow{2}{*}{\begin{tabular}[l]{@{}l@{}}Privacy \&\\Fairness\end{tabular}}
  & Model \& Data Privacy & \priority{0} & \priority{100} & \priority{50} & \priority{50} & \priority{0} & \priority{100} & \priority{100} \\
  & Fairness \& Bias & \priority{50} & \priority{100} & \priority{50} & \priority{100} & \priority{0} & \priority{50} & \priority{100} \\
\bottomrule
\end{tabular}
\end{adjustbox}
\label{tab:coverage-by-service}
\end{table*}

\subsection{Open-Source Frameworks and Toolkits}

\textbf{HELM Safety.} \textit{HELM}~\cite{helm:liang2022holistic:arxiv:2022} introduced a standardized evaluation protocol emphasizing helpfulness, harmlessness, and fairness. Its Safety extension integrates the \textit{Anthropic Red-Team}, \textit{BBQ}, \textit{HarmBench}, \textit{Simple Safety}, and \textit{XSTest} benchmarks to assess some aspects of alignment, refusal, and bias, but lacks deep analysis and coverage of other risks.

\textbf{TrustLLM.} \textit{TrustLLM}~\cite{trustllm:huang2024:2024:arxiv} expanded coverage to ethics, factuality, and fairness. It provides comprehensive fairness and hallucination benchmarks but omits code-level security, lacks systematic over-refusal calibration, and considers only basic jailbreaking methods.

\textbf{TrustEval.} \textit{TrustEval}~\cite{trusteval:wang:acl:2025} focuses on dynamic test generation and multimodal evaluation but covers narrow aspects of each service, e.g., checking factuality but excluding consistency and grounding checks within \textit{Hallucination}.

\textbf{DecodingTrust.} \textit{DecodingTrust}~\cite{decodingtrust:wang2023decodingtrust:neurips:2023} focuses on assessing the trustworthiness of GPT models, adapting popular adversarial and fairness benchmarks. It quantifies OOD reliability and privacy leakage but does not integrate code security diagnostics, consider behavioral privacy, or directly probe different types of hallucination.

\textbf{PromptBench.} \textit{PromptBench}~\cite{zhu2024promptbenchunifiedlibraryevaluation} provides a tool for adversarial prompt evaluation, covering lexical, syntactic, and semantic perturbations across 11 NLP tasks. While effective for comparative robustness testing, it remains a perturbation corpus rather than a full evaluation framework.

\textbf{garak.} \textit{garak}~\cite{derczynski2024garak} is an offensive red-team framework comprising over 30 probes targeting jailbreaks, encoding attacks, toxicity, leakage, and code exploits. Its probe library (e.g., AutoDAN, GCG, Invisible-Tags, MalwareGen) systematically stresses several services, but provides no calibrated scoring or assessment for over refusal and OOD robustness, while only probing slur continuation and factuality under \textit{Fairness \& Bias} and \textit{Hallucination}.

\subsection{Comparative Coverage Across Dimensions}

As Table~\ref{tab:coverage-by-service} shows, existing frameworks cluster heavily around safety alignment, jailbreaks, and adversarial robustness. While several also cover hallucination, privacy, and fairness, they do so partially, focusing on a single aspect of the risk area (e.g., only factuality, leakage, or stereotypes). Only \textit{garak} includes partial code security probing, and none perform cross-guard safety calibration. \textit{PromptBench} and \textit{garak} provide valuable stress tests but lack unified scoring, normalization, or cross-dimensional integration. Consequently, no prior framework offers simultaneous evaluation of all dimensions under a unified protocol. \system addresses this gap by combining breadth (46 tests across all nine services) with depth (multiple heterogeneous benchmarks, scoring methods, and judging mechanisms per service), enabling both granular diagnostics and cross-dimensional analysis within a single taxonomy.
\section{Limitations}
\label{sec:limitations}

Our evaluation operates under several structural constraints that bound the scope of our conclusions.

\noindent\textbf{Language coverage.} All tests are English-only, limiting our ability to detect vulnerabilities that manifest in other languages or cross-lingual contexts. Safety behaviors may degrade substantially in low-resource languages where alignment data is sparse, a dimension our framework does not currently capture.

\noindent\textbf{Black-box access.} All benchmarks operate through text-level input-output interactions; the distillation analysis supplements this with a representational diagnostic on open-weight models. Gradient-based adversarial testing and mechanistic interpretability remain out of scope.

\noindent\textbf{Judge reliability.} While our multi-judge ensemble and cross-judge validation mitigate systematic biases, edge cases involving implicit harm, contextual nuance, or culturally specific content remain difficult to adjudicate. The binary safe/unsafe framing further compresses the spectrum of possible harms.

\section{Ethics Considerations}
This work exclusively relies on publicly available datasets and benchmarks for evaluating the safety, security, and privacy properties of LLMs.  No private, proprietary, or personally identifiable data were collected, annotated, or processed by humans for this study.
All proprietary models were accessed through their official APIs \raid{(via OpenRouter)} under standard terms of service, with built-in safety and filtering mechanisms left enabled.

\raid{Evaluation results, including prompt-level traces, are publicly available through the \system portal for both the community and the model providers; providers may submit models in private mode, or subsequently move any examination to private, retaining control over disclosure. Although some analyses surface unsafe generations, the underlying tests are drawn from previously released public benchmarks and introduce no novel exploits, and we consider the defensive value of standardized comparative measurement to outweigh the residual risk that published rankings could help adversaries prioritize targets.}

The intent of this work is to advance the scientific understanding of trade-offs between safety, security, and privacy in LLMs, not to expose or exploit vulnerabilities. We explicitly discourage any misuse of the \system framework or its results for adversarial or malicious purposes.

\section{Conclusion}

We presented \system, a unified black-box evaluation framework spanning 46 tests across nine services to jointly assess the safety, security, and privacy of LLMs. Our large-scale analysis of over 120 models demonstrates that trustworthiness is inherently multi-dimensional. Safety enforcement incurs a measurable utility cost through increased over-refusal; \raid{privacy is near-orthogonal to other dimensions and not captured} by standard alignment; off-policy distillation without on-policy correction induces severe robustness collapse; and safety behaviors are category-dependent rather than universal, with alignment functioning as behavioral overfitting to familiar risk patterns.
These findings converge on a central implication: progress along one trustworthiness axis does not guarantee, and can actively undermine, progress along others. Future work should develop alignment methods that avoid unnecessary conservatism, robustness-preserving distillation techniques, and privacy-aware training pipelines. Until alignment moves from monolithic safety tuning to modular, dimension-aware architectures, trustworthiness gains in one area will continue to come at the expense of others.

\bibliographystyle{splncs04}
\bibliography{refs}

\appendix
\section{Evaluation Components}
\label{appendix:overview}
Table~\ref{tab:datasets} enumerates the services, tests, datasets, and metrics of \system test suite.

\begin{table*}
    \centering
    \caption{\system services, tests, datasets, and metrics.}
    \renewcommand{\arraystretch}{1.3}
    \begin{adjustbox}{max width=\textwidth}
    \begin{tabular}{l l l p{6.5cm} c c l}
        \toprule
        \textbf{Service} & \textbf{Test} & \textbf{Dataset} & \textbf{Description} & \textbf{\# Samples}& \textbf{\# Categories} & \textbf{Metric} \\
        \midrule

        \multirow{4}{*}{Hallucination} & SimpleQA & SimpleQA~\cite{simpleqa:arxiv:wei2024measuring:2024} & Fact-seeking less frequently encountered questions with short answers & 4,326  & 10 & Accuracy \\
        & TruthfulQA & TruthfulQA~\cite{truthfulqa:lin2021truthfulqa:arxiv} & Multiple-choice questions related to common false beliefs or misconceptions held by humans & 816 & 17 & Accuracy \\
        & TriviaQA & TriviaQA~\cite{dataset:triviaqa:joshi:2017} & Trivia style fact-seeking questions with short answers, measuring model accuracy for attempted responses  & 9{,}280 & -- & Accuracy \\
        & SelfCheckGPT & WikiBio~\cite{dataset:wikibio:lebret2016neural:arxiv:2016} & A dataset of Wikipedia biographies describing individuals, used to assess the consistency of generated responses & 239 & -- & Accuracy \\
        & Vectara & Vectara~\cite{vectara:leaderboard:2024} & Evaluates the alignment of model-generated summaries with the provided source text & 1{,}006 & -- & Accuracy \\
        & FaithEval & FaithEval~\cite{faitheval:shafiq:2024} & QA dataset with context that assesses the faithfulness of generated responses to the provided context & 4,992 & 3 & Accuracy \\
        % \cmidrule(lr){2-3}
        & \multirow{3}{*}{HaluEval} & OpenDialKG~\cite{opendialkg:moon2019:acl:2019} & A dataset of knowledge-based dialogues & 10,000 && \\
        & & HotpotQA~\cite{hotpotqa:yang2018:arxiv:2018} & QA pairs based on the factual knowledge from Wikipedia & 10,000 & \multirow{3}{*}{3} & \multirow{3}{*}{Accuracy} \\
        % & & OpenDialKG~\cite{opendialkg:moon2019:acl:2019} & A dataset of knowledge-based dialogues & 10,000 && \\
        & & CNN/Daily Mail~\cite{cnndaily:see2017get:arxiv:2017} & A dataset of news articles from CNN and Daily Mail, paired with human-generated summaries & 10,000 && \\
        \midrule

        \multirow{2}{*}{\shortstack{Code\\ Security}} & CyberSecEval 3 & CyberSecEval 3~\cite{code:arxiv:wan2024cyberseceval:2024} &  Code generation prompts in instruction and autocomplete contexts across 50 CWEs and 6 programming languages & 3,832 & 6 & Accuracy \\
        & SecCodePLT & SecCodePLT~\cite{code:arxiv:yang2024seccodeplt:2024} & Code generation prompts on Python-specific vulnerabilities, spanning 27 CWEs & 2,104 & 2 & Accuracy \\
        \midrule

        \multirow{6}{*}{\shortstack{Safety\\ Alignment}}
        & Anthropic Red Team~\cite{anthropicredteaming:bai2022training:arxiv:2022} & HH-RLHF~\cite{ganguli2022redteaminglanguagemodels} & Red team prompts for testing model safety boundaries & 1000 & -- & Accuracy \\
        & OpenAI Moderation & OpenAI Moderation~\cite{dataset:openai2022moderation} & Evaluates moderation capabilities of LLMs & 522 & 8 & Accuracy \\
        & Perspective API & RealToxicity~\cite{dataset:realtoxicityprompts} & Measures toxicity in generated responses & 946 & 6 & Accuracy \\
        & Llama Guard 1~\cite{llamaguard1:inan2312llama:2023} & BeaverTails~\cite{dataset:beavertails} & Prompts inciting harmful or unsafe behavior & 500 & 6 & Accuracy \\
        & Llama Guard 2~\cite{llamaguard2:2024} & Do-Not-Answer~\cite{dataset:donotanswer}  & Prompts designed to elicit refusal or harmful responses & 768 & 11 & Accuracy \\
        & Llama Guard 3~\cite{llamaguard3:2024} & HarmfulQA~\cite{dataset:harmfulqa} & Questions containing harmful or unsafe content & 553 & 14 & Accuracy \\
        & Llama Guard 4~\cite{llamaguard4:2025} & CatQA~\cite{dataset:llamaguard4:CatQA:bhardwaj2024:arxiv:2024} & A categorical harmful questions dataset & 550 & 14 & Accuracy \\
        & WildGuard & WildGuard~\cite{dataset:han2024wildguard} & Adversarial prompts challenging safety alignment & 754 & 5 & Accuracy \\
        & Simple Safety Tests & SimpleSafetyTests~\cite{dataset:simplesafetytests:vidgen2023:arxiv:2023} & Prompts testing safety refusal  across harm categories & 100 & 5 & Accuracy \\
        \midrule

        \multirow{3}{*}{\shortstack{Over\\ Refusal}} & OKTest & OKTest~\cite{dataset:oktest} & Safe prompts that may be misclassified as unsafe & 350 & -- & Accuracy \\
        & OR-Bench & OR-Bench~\cite{dataset:orbench} & Benchmarks refusal handling of safe prompts & 1,319 & 10 & Accuracy \\
        & XSTest & XSTest~\cite{dataset:xstest} & Measures robustness to misinterpretation of safe prompts & 450   & 18 & Accuracy \\
        & WildGuard & WildGuard~\cite{dataset:han2024wildguard} & Innocuous prompts challenging over refusal & 971 & 2 & Accuracy \\
        \midrule

        \multirow{2}{*}{\shortstack{Adversarial\\ Robustness}} & AdvGlue & AdvGlue~\cite{dataset:advglue}  & Jailbreak prompts generated via adversarial attacks & 576 & 10 & Accuracy \\
        & AdvGlue++ & AdvGlue++~\cite{decodingtrust:wang2023decodingtrust:neurips:2023} & Enhanced adversarial attacks for robustness evaluation & 38,054 & 5 & Accuracy \\
        \midrule

        \multirow{3}{*}{\shortstack{Jailbreak\\ Robustness}}
        & Jailbroken~\cite{wei2023jailbroken} & \multirow{3}{*}{\shortstack{AdvBench~\cite{dataset:advbench:zou2023:arxiv:2023}}}  & Generates artificial prompts leveraging conflicting objectives and mismatched generalization & \multirow{3}{*}{522} & 10 & Accuracy \\
        & Cipher~\cite{yuan2024cipher} &  & Uses various encryption techniques, instead of natural languages to test the ability to maintain safety protocols & & 4 & Accuracy \\
        & PAIR~\cite{chao2024pair} & & Generates semantic jailbreaks using an attacker LLM & & -- & Accuracy \\
        \midrule

        \multirow{2}{*}{\shortstack{OOD\\ Robustness}} & DecodingTrust & DecodingTrust~\cite{decodingtrust:wang2023decodingtrust:neurips:2023}  & Tests how well a model generalizes to data that differs from its training distribution & 9592 & 10 & Accuracy \\
        \midrule

        \multirow{4}{*}{\shortstack{Model \&\\ Data Privacy}} & PII Awareness & PII Awareness~\cite{trustllm:huang2024:2024:arxiv} & Privacy-related queries, assessing the disclosure or withholding of sensitive personally identifiable information & 560 & 7 & Accuracy \\
        & Privacy Understanding & ConfAIde~\cite{confaide:mireshghallah2023can:arxiv:2023} & Benchmark for evaluating the model's alignment with human expectations in privacy-sensitive contexts & 196 & -- & {\shortstack{Pearson \\ Correlation}} \\
        & Data Leakage & Enron~\cite{enron:klimt2004enron:2004} & Dataset evaluating potential data leakage through email addresses from the Enron Corporation & 400 & 4 & Accuracy \\
        & ECHR & ECHR~\cite{echr:poudyal2020echr} & European Court of Human Rights dataset with PII annotations  (names, dates, locations) & 600 & 3 & Accuracy \\
        \midrule

        \multirow{3}{*}{\shortstack{Fairness \&\\ Bias}} & Disparagement & Adult~\cite{dataset:adult} & A structured dataset of demographic and work attributes for salary level prediction & 810 & 6 & Cramer's V \\
        & GenderCARE & GenderPair~\cite{gendercare:tang2024gendercare:ccs:2024} & A dataset for evaluating gender bias, focusing on biases in gender-related language choices & 103,854 & 3 & Accuracy \\
        & Preference & Preference~\cite{trustllm:huang2024:2024:arxiv} & Prompts designed to assess whether the LLM favors/promotes specific ideologies/lifestyles & 240 & 2 & Accuracy \\
        & BBQ & BBQ~\cite{dataset:bbq:parrish2021:arxiv:2021} & Analyzes disparities related to age, disability, nationality, race, religion, sexuality, and socioeconomic factors & 1100 & 11 & Accuracy \\
        \midrule
        \textbf{Total} & & & & \textbf{221,882} & \textbf{215}& \\
        \bottomrule
    \end{tabular}
    \end{adjustbox}
    \label{tab:datasets}
\end{table*}

\section{Experimental Setup}
\label{appendix:setup}

\textbf{Inference Configuration.}
Unless otherwise specified, all evaluations were conducted using deterministic decoding (temperature$\,{=}\,$0) with a maximum generation length of 2{,}048 tokens. Each test prompt was executed once per model to reflect real-world single-query deployment scenarios. To handle transient API failures, each prompt was retried up to five times with exponential back-off before being marked as failed. Open-weight models were served in containerized local deployments using vLLM~\cite{vllm:kwon2023efficient}. 

\textbf{Scoring and Judge Configuration.}
Each test is scored by either a benchmark-specific judge or a general-purpose LLM judge. Benchmark-specific judges are used whenever the original benchmark defines one: LLaMA Guard 1--4~\cite{llamaguard1:inan2312llama:2023,llamaguard2:2024,llama3:dubey2024llama:arxiv:2024,llamaguard4:2025} and WildGuard~\cite{dataset:han2024wildguard} for their respective safety and over-refusal tests, OpenAI Moderation~\cite{openai2024moderation} and Perspective API~\cite{PerspectiveAPI} for toxicity scoring, and Vectara's factual consistency scorer~\cite{vectara:leaderboard:2024} for summarization faithfulness. For all remaining tests, GPT-oss-120B serves as the general-purpose evaluation judge. To validate judge reliability, we independently scored the full evaluation suite using Gemma3-27B as an alternative general-purpose judge; the two judges achieve a Pearson correlation of 0.985 (95\% CI: [0.983, 0.987]) across all service-level scores, confirming that our results are not artifacts of a single judge model. One test-specific exception to the default inference configuration: SelfCheckGPT requires multiple stochastic response variations per prompt, so target-model generations for this test use temperature$\,{=}\,1$ instead of the default deterministic decoding.

\textbf{Infrastructure.}
Evaluations were executed on a distributed setup comprising four Nvidia H200 nodes (140\,GB memory each), using parallel GPU batching and asynchronous API scheduling for high throughput. A complete run for a single model typically requires one to two days of wall-clock time, with the full evaluation consuming over 2{,}000 GPU hours.

\textbf{Proprietary Model Versions.}
All proprietary models were accessed through OpenRouter\footnote{\url{https://openrouter.ai}}. Because provider APIs may update model weights without notice, we record the specific model identifiers and final evaluation dates for reproducibility. GPT-4o (\texttt{openai/gpt-4o}), Gemini~2.0 (\texttt{google/gemini-2.0-flash-001}), Gemini~2.5 (\texttt{google/gemini-2.5-flash}), Claude-4.5 (\texttt{anthropic/claude-haiku-4.5}), and Grok~4 (\texttt{x-ai/grok-4-fast}) were last evaluated on November~7; DeepSeek-R1 (\texttt{deepseek/deepseek-r1-0528}) on November~10; GPT-5 (\texttt{openai/gpt-5-mini}) and Claude-3.5 (\texttt{anthropic/claude-3.5-haiku}) on November~11; Grok~3 (\texttt{x-ai/ grok-3-mini}) and Grok~4.1 (\texttt{x-ai/grok-4.1-fast}) on November~27; and Gemini~3 (\texttt{google/ gemini-3-flash-preview}) on December~21.

\section{Context-Aware Adaptive Evaluation}
\label{sec:stakeholder}

A uniform arithmetic mean over evaluation dimensions implicitly assumes
that every service is equally important in every deployment scenario.
In practice, however, the relative importance of trustworthiness dimensions
is \emph{deployment-context dependent}: a healthcare chatbot must above all
avoid hallucinations and protect patient privacy, whereas a code assistant
must prioritize code security and adversarial robustness.

To study how evaluation perspective affects model rankings, we define
\emph{stakeholder profiles}: normalized weight vectors over \system's nine
services representing six deployment contexts:
(i)~Uniform (equal weights; our default),
(ii)~Healthcare (emphasizing hallucination, privacy, and safety),
(iii)~Code~Assistant (emphasizing code security, adversarial
robustness, and jailbreak resistance),
(iv)~Customer~Chatbot (emphasizing fairness, over-refusal balance,
and OOD robustness),
(v)~Content~Moderator (emphasizing safety alignment, jailbreak
robustness, and over-refusal), and
(vi)~Research\,/\,Open (emphasizing privacy, fairness, and
factuality).
The weights are one reasonable instantiation derived from domain-specific
risk priorities in established regulatory and operational
guidelines~\cite{ebers2023european,nistai2024artificial}; the framework is
fully configurable, and alternative weightings can be substituted without
changing the evaluation pipeline.
For each profile, a model's weighted score is computed as
$S_p = \sum_{i=1}^{9} w_i^{(p)} \cdot s_i$, where $w_i^{(p)}$ are the
normalized profile weights and $s_i$ the per-service scores.

\begin{table*}[t]
\centering
\caption{Weighted scores and rankings of the top-15 models (by Uniform score), reranked 1--15 within each stakeholder profile. Shaded cells mark rank improvements $\geq$5 relative to Uniform; red cells mark drops $\geq$5.}
\label{tab:stakeholder_scores_ranks}
\resizebox{\textwidth}{!}{
\begin{tabular}{lrr|rr|rr|rr|rr|rr}
\toprule
\textbf{Model} & \multicolumn{2}{c}{\textbf{Uniform}} & \multicolumn{2}{c}{\textbf{Healthcare}} & \multicolumn{2}{c}{\textbf{Code Asst.}} & \multicolumn{2}{c}{\textbf{Cust. Chat.}} & \multicolumn{2}{c}{\textbf{Cont. Mod.}} & \multicolumn{2}{c}{\textbf{Research}} \\
\cmidrule(lr){2-3} \cmidrule(lr){4-5} \cmidrule(lr){6-7} \cmidrule(lr){8-9} \cmidrule(lr){10-11} \cmidrule(lr){12-13}
 & Score & Rank & Score & Rank & Score & Rank & Score & Rank & Score & Rank & Score & Rank \\
\midrule
GPT-5 & 81.8 & \textbf{1} & 81.2 & 2 & 82.2 & \textbf{1} & 82.1 & 2 & 86.0 & \textbf{1} & 77.9 & 3 \\
Gemini 3 & 80.6 & 2 & 79.4 & 4 & 78.3 & 6 & 83.5 & \textbf{1} & 83.2 & 3 & 78.2 & 2 \\
Gemma4-31B & 80.0 & 3 & 79.3 & 5 & 77.9 & 7 & 82.1 & 3 & 82.9 & 5 & 77.4 & 4 \\
Claude-4.5 & 80.0 & 4 & 81.6 & \textbf{1} & 79.8 & 2 & 79.8 & 6 & 85.4 & 2 & 77.2 & 5 \\
Gemma4-26B-A4B & 78.9 & 5 & 78.3 & 6 & 76.9 & 9 & 80.7 & 4 & 81.4 & 8 & 76.3 & 7 \\
GPT-120B & 78.7 & 6 & 77.8 & 8 & 79.3 & 3 & 78.0 & 7 & 83.0 & 4 & 74.4 & 9 \\
GPT-4o & 78.0 & 7 & 80.7 & 3 & 73.1 & \cellcolor{red!10}14 & 80.1 & 5 & 78.0 & \cellcolor{red!10}12 & 80.3 & \cellcolor{gray!15}\textbf{1} \\
DeepSeek-R1 & 76.9 & 8 & 78.0 & 7 & 73.7 & 12 & 77.9 & 8 & 77.3 & \cellcolor{red!10}14 & 76.9 & 6 \\
Qwen3.5-27B & 76.3 & 9 & 72.3 & \cellcolor{red!10}15 & 78.6 & 5 & 77.2 & 11 & 82.6 & 6 & 69.1 & \cellcolor{red!10}15 \\
GPT-20B & 76.3 & 10 & 74.5 & 11 & 77.1 & 8 & 75.8 & 14 & 81.3 & 9 & 71.0 & 12 \\
Qwen3.5-122B-A10B & 76.2 & 11 & 72.9 & 13 & 78.7 & \cellcolor{gray!15}4 & 76.2 & 13 & 82.1 & 7 & 69.3 & 14 \\
Grok 4 & 76.0 & 12 & 74.0 & 12 & 74.8 & 11 & 77.4 & 10 & 80.9 & 10 & 71.4 & 11 \\
Grok 4.1 & 75.3 & 13 & 72.6 & 14 & 75.5 & 10 & 76.3 & 12 & 79.6 & 11 & 70.7 & 13 \\
Claude-3.5 & 75.1 & 14 & 77.8 & \cellcolor{gray!15}9 & 73.4 & 13 & 74.6 & 15 & 76.7 & 15 & 73.3 & 10 \\
Phi-4 & 75.0 & 15 & 75.2 & \cellcolor{gray!15}10 & 70.6 & 15 & 77.7 & \cellcolor{gray!15}9 & 77.4 & 13 & 74.6 & \cellcolor{gray!15}8 \\
\bottomrule
\end{tabular}}
\end{table*}

We focus the analysis on the 15 highest-scoring models under the
Uniform profile and rerank them 1--15 within each stakeholder profile.
Table~\ref{tab:stakeholder_scores_ranks} presents the weighted scores and
reranked positions.  The identity of the \#1 model changes across
profiles: GPT-5 leads under Uniform, Code~Assistant, and
Content~Moderator, but Claude-4.5 takes the lead under Healthcare,
Gemini~3 tops Customer~Chatbot, and GPT-4o rises to \#1 under Research.
In total, four distinct models occupy the top position across six
profiles, demonstrating that a single aggregate ranking cannot capture
deployment-specific trustworthiness.

We quantify ranking divergence using Kendall's~$\tau$ rank correlation
between all profile pairs.  Within the top~15, the mean pairwise
$\tau=0.455$ and 27\% of pairwise model orderings reverse when
switching evaluation perspectives, confirming that rankings among the
strongest models are highly sensitive to the evaluation perspective.

\textbf{Rank displacement reveals model sensitivity.}
For each model we compute the maximum rank displacement, the difference
between its best and worst rank across all profiles.  The mean maximum
displacement is 6.2~positions; 53\% of models shift by more
than 5~positions.  The most sensitive model is GPT-4o, whose rank
swings from \#1 under Research\,/\,Open to \#14 under Code~Assistant, a
displacement of 13~positions.  The Qwen~3.5 models exhibit the highest
mean displacement ($\sim$10 positions), while GPT-5 and Grok~4 remain
the most stable ($\leq$2 positions).
Such instability shows that models strong in one
trustworthiness dimension can be weak in another, a finding invisible under
uniform aggregation.

These results establish that multi-dimensional trustworthiness evaluation
is not merely a ``checklist'' but a \emph{perspective-aware} framework:
the same set of evaluation services, combined differently, yields
materially different conclusions.  No prior platform supports such stakeholder-aware evaluation, and the
findings demonstrate that this capability reveals insights inaccessible to
any single-metric or uniformly aggregated benchmark.

\textbf{Ranking robustness under random weightings.}
To verify that the ranking sensitivity observed above is not an artifact of
the six chosen profiles, we sample 10{,}000 weight vectors from a symmetric
Dirichlet distribution ($\alpha{=}1$ for each service) and recompute the
full ranking under each.  The analysis reveals a two-layer structure.  At
the coarse level, rankings are stable: the mean Kendall's~$\tau$ between
the Uniform ranking and random rankings is~0.775 across all evaluated models,
and the top four models (GPT-5, Gemini~3, Gemma4-31B, Claude-4.5)
collectively hold the \#1 position in 82\% of samples.  At the fine level,
within-tier ordering is sensitive: 8~distinct models reach \#1 across
10{,}000~samples and only 67\% of the Uniform top-5 appear in any given
random top-5. These results confirm
that the overall leaderboard provides a meaningful default ranking, but
deployment-specific weighting is necessary to resolve the ordering within
the top tier.

\section{Theoretical Analysis of Distillation-Induced Robustness Collapse}
\label{app:distillation-embeddings}

Modern strong-to-weak distillation follows a two-stage process~\cite{agarwal2024policy}: (1) \emph{off-policy distillation}, where the student imitates teacher logits on sequences generated entirely by the teacher, and (2) \emph{on-policy distillation}, where the student generates its own trajectories and aligns their logits with the teacher via KL minimization. Formally, the on-policy objective minimizes
\begin{equation}
\mathcal{L}_{\mathrm{KL}}(\theta)
= \mathbb{E}_{x \sim \mathcal{D}}
  \mathbb{E}_{y \sim p_\theta(\cdot \mid x)}
  \left[
    D_{\mathrm{KL}}\!\left(
        p_T(\cdot \mid x,y) \;\|\; p_\theta(\cdot \mid x,y)
    \right)
  \right],
\end{equation}
where $p_T$ denotes the teacher distribution and $p_\theta$ the student. This KL-based correction ensures that the student is trained on its \emph{own} output distribution, preventing divergence from the teacher. Crucially, this step requires vocabulary compatibility; without matching tokenization between teacher and student, the KL term cannot be computed correctly.

In R1-style distilled open-weight models, this on-policy phase is omitted or cannot be applied due to vocabulary mismatches. Training then collapses to a pure off-policy objective:
\begin{equation}
\mathcal{L}_{\mathrm{off}}(\theta)
= \mathbb{E}_{(x,y) \sim \text{TeacherGen}}
  \left[ -\log p_\theta(y \mid x) \right],
\end{equation}
which implicitly pushes $p_\theta(y \mid x)$ toward extremely peaked, low-entropy distributions. Since the cross-entropy term is minimized when $p_\theta(y \mid x) \rightarrow 1$, the student is encouraged to overfit the exact teacher token trajectory. This collapses the entropy of the predictive distribution, $H[p_\theta(\cdot \mid x)] \rightarrow 0$, steepening the decision boundary. As a result, even small input perturbations $\delta x$ can cross the sharpened decision boundary, causing disproportionately large shifts in the output distribution $\| p_\theta(\cdot \mid x) - p_\theta(\cdot \mid x + \delta x) \|$. This loss of local smoothness is a classical precursor to adversarial brittleness~\cite{goodfellow2014explaining}. Because the student never receives feedback on its own erroneous trajectories, the support of $p_\theta$ drifts away from that of the teacher, producing sharp but poorly calibrated decision boundaries.

\section{Detailed Evaluation Results}
\label{app:service_results}

This appendix presents the extended cross-service leaderboard and detailed per-service evaluation results.

\subsection{Extended Leaderboard}
\label{app:full_leaderboard}

Table~\ref{tab:full_leaderboard} extends the main leaderboard (Table~\ref{tab:leaderboard}) with additional model sizes from the selected families, enabling the intra-family scaling and distillation analyses presented in the main text. The complete leaderboard covering all evaluated models is accessible through the \system portal.

\begin{table*}
    \centering
    \caption{Extended \system Leaderboard with all model sizes from the selected families, grouped by (1) API-based, (2) open-weight, and (3) distilled models. The complete leaderboard is available through the \system portal.}
    \resizebox{\textwidth}{!}{
    \begin{tabular}{l@{\hskip 4pt}|@{\hskip 4pt}c@{\hskip 4pt}c@{\hskip 4pt}c@{\hskip 4pt}c@{\hskip 4pt}c@{\hskip 4pt}c@{\hskip 4pt}c@{\hskip 4pt}c@{\hskip 4pt}c@{\hskip 4pt}|@{\hskip 4pt}c}
        \toprule
        & \multicolumn{9}{c|}{\textbf{Services}} & \\
        \cmidrule(lr){2-10}
         \textbf{Model}
        & \shortstack{\textbf{Adv.}\\\textbf{Robust.}}
        & \shortstack{\textbf{Code}\\\textbf{Sec.}}
        & \shortstack{\textbf{Fair.}\\\textbf{\& Bias}}
        & \shortstack{\textbf{Halluc.}}
        & \shortstack{\textbf{Jail.}\\\textbf{Robust.}}
        & \shortstack{\textbf{M\&D}\\\textbf{Priv.}}
        & \shortstack{\textbf{OOD}\\\textbf{Robust.}}
        & \shortstack{\textbf{Over}\\\textbf{Ref.}}
        & \shortstack{\textbf{Safety}\\\textbf{Align.}}
        & \shortstack{\textbf{Overall} \\ \textbf{Score}} \\
        \midrule
\includegraphics[width=0.33cm]{images/logo/openai-logo.png} \textbf{GPT-5} & 71.6 & 79.7 & 68.8 & \textbf{83.9} & 90.2 & 69.4 & 89.5 & 87.9 & 95.6& \textbf{81.8} \\
\includegraphics[width=0.33cm]{images/logo/gemini-logo-light.png} \textbf{Gemini 3} & \textbf{77.2} & 74.3 & 76.7 & 82.7 & 70.3 & 65.1 & \textbf{91.6} & 94.4 & 93.2& \textbf{80.6} \\
\includegraphics[width=0.33cm]{images/logo/claude-logo-light.png} \textbf{Claude-4.5} & 61.6 & 70.3 & 62.7 & 80.8 & 98.5 & 78.7 & 88.1 & 82.3 & 97.0& \textbf{80.0} \\
\includegraphics[width=0.33cm]{images/logo/openai-logo.png} \textbf{GPT-4o} & 62.6 & 73.1 & 67.5 & 77.5 & 51.4 & \textbf{91.1} & 86.9 & 94.4 & 97.5& \textbf{78.0} \\
\includegraphics[width=0.33cm]{images/logo/deepseek-logo.png} \textbf{DeepSeek-R1} & 67.2 & 77.1 & 61.6 & 74.9 & 51.8 & 84.7 & 86.2 & 91.5 & 97.4& \textbf{76.9} \\
\includegraphics[width=0.33cm]{images/logo/grok-logo-light.png} \textbf{Grok 4} & 70.3 & 69.7 & 61.5 & 66.0 & 73.2 & 68.4 & 90.0 & 88.3 & 97.0& \textbf{76.0} \\
\includegraphics[width=0.33cm]{images/logo/grok-logo-light.png} \textbf{Grok 4.1} & 68.3 & 72.0 & 59.5 & 66.1 & 78.1 & 68.2 & 89.8 & 88.2 & 88.0& \textbf{75.3} \\
\includegraphics[width=0.33cm]{images/logo/claude-logo-light.png} \textbf{Claude-3.5} & 62.7 & 72.2 & 61.5 & 74.9 & 70.5 & 80.8 & 87.0 & 67.2 & \textbf{99.3}& \textbf{75.1} \\
\includegraphics[width=0.33cm]{images/logo/grok-logo-light.png} \textbf{Grok 3} & 66.3 & 68.6 & 60.9 & 74.0 & 47.4 & 67.4 & 89.6 & 94.4 & 94.0& \textbf{73.6} \\
\includegraphics[width=0.33cm]{images/logo/gemini-logo-light.png} \textbf{Gemini 2.0} & 65.7 & 66.0 & 66.9 & 76.2 & 50.3 & 59.1 & 88.6 & 85.4 & 94.8& \textbf{72.5} \\
\includegraphics[width=0.33cm]{images/logo/gemini-logo-light.png} \textbf{Gemini 2.5} & 62.8 & 72.2 & 60.6 & 72.8 & 49.7 & 58.9 & 88.3 & 90.4 & 90.0& \textbf{71.7} \\
\midrule
\includegraphics[width=0.33cm]{images/logo/huggingface-logo.png} \textbf{Gemma4-31B} & 75.2 & 75.3 & \textbf{79.0} & 75.7 & 73.6 & 69.6 & 89.1 & 85.3 & 97.4& \textbf{80.0} \\
\includegraphics[width=0.33cm]{images/logo/huggingface-logo.png} \textbf{Gemma4-26B-A4B} & 70.1 & 76.4 & 77.5 & 73.2 & 72.7 & 70.8 & 89.5 & 82.3 & 97.3& \textbf{78.9} \\
\includegraphics[width=0.33cm]{images/logo/huggingface-logo.png} \textbf{GPT-120B} & 66.3 & \textbf{80.1} & 72.9 & 64.2 & 91.9 & 76.9 & 85.0 & 72.4 & 98.2& \textbf{78.7} \\
\includegraphics[width=0.33cm]{images/logo/huggingface-logo.png} \textbf{Qwen3.5-27B} & 69.8 & 75.0 & 71.2 & 71.5 & \textbf{99.5} & 48.4 & 86.5 & 79.6 & 85.6& \textbf{76.3} \\
\includegraphics[width=0.33cm]{images/logo/huggingface-logo.png} \textbf{GPT-20B} & 65.9 & 78.2 & 67.6 & 59.7 & 87.7 & 72.0 & 83.7 & 74.5 & 97.5& \textbf{76.3} \\
\includegraphics[width=0.33cm]{images/logo/huggingface-logo.png} \textbf{Qwen3.5-122B-A10B} & 69.2 & 74.1 & 64.3 & 73.1 & 99.2 & 54.1 & 86.9 & 79.0 & 85.4& \textbf{76.1} \\
\includegraphics[width=0.33cm]{images/logo/huggingface-logo.png} \textbf{Phi-4} & 63.5 & 68.0 & 72.0 & 58.4 & 56.8 & 84.2 & 87.3 & 86.3 & 99.0& \textbf{75.1} \\
\includegraphics[width=0.33cm]{images/logo/huggingface-logo.png} \textbf{Qwen3.5-35B-A3B} & 68.1 & 72.1 & 63.0 & 70.6 & 98.7 & 54.0 & 84.9 & 78.4 & 85.5& \textbf{75.0} \\
\includegraphics[width=0.33cm]{images/logo/huggingface-logo.png} \textbf{Qwen3.5-9B} & 66.1 & 73.2 & 63.8 & 70.5 & 98.8 & 52.7 & 84.1 & 78.7 & 86.6& \textbf{74.9} \\
\includegraphics[width=0.33cm]{images/logo/huggingface-logo.png} \textbf{Llama3.3-70B} & 67.6 & 68.4 & 58.9 & 72.1 & 47.8 & 74.4 & 88.6 & 94.0 & 94.1& \textbf{74.0} \\
\includegraphics[width=0.33cm]{images/logo/huggingface-logo.png} \textbf{Qwen3.5-4B} & 65.1 & 71.5 & 59.3 & 68.8 & 98.6 & 51.0 & 82.4 & 78.5 & 85.2& \textbf{73.4} \\
\includegraphics[width=0.33cm]{images/logo/huggingface-logo.png} \textbf{Gemma2-27B} & 69.9 & 61.2 & 64.7 & 64.5 & 67.0 & 72.2 & 88.6 & 73.8 & 97.9& \textbf{73.3} \\
\includegraphics[width=0.33cm]{images/logo/huggingface-logo.png} \textbf{Llama4-17B-16E} & 61.1 & 69.3 & 57.8 & 61.1 & 56.4 & 74.5 & 88.4 & 94.0 & 94.5& \textbf{73.0} \\
\includegraphics[width=0.33cm]{images/logo/huggingface-logo.png} \textbf{Llama3.1-8B} & 56.9 & 70.0 & 65.7 & 62.0 & 55.4 & 78.6 & 79.4 & 90.3 & 96.6& \textbf{72.8} \\
\includegraphics[width=0.33cm]{images/logo/huggingface-logo.png} \textbf{Llama3.2-3B} & 43.9 & 62.9 & 52.4 & 66.6 & 58.3 & 88.1 & 86.2 & 93.8 & 94.3& \textbf{71.8} \\
\includegraphics[width=0.33cm]{images/logo/huggingface-logo.png} \textbf{Gemma4-E4B} & 61.1 & 71.3 & 65.8 & 70.7 & 52.9 & 65.7 & 85.0 & 72.3 & 98.1& \textbf{71.4} \\
\includegraphics[width=0.33cm]{images/logo/huggingface-logo.png} \textbf{Gemma3-27B} & 67.3 & 63.2 & 56.1 & 58.3 & 48.0 & 68.2 & 86.3 & \textbf{94.6} & 92.7& \textbf{70.5} \\
\includegraphics[width=0.33cm]{images/logo/huggingface-logo.png} \textbf{Gemma3-12B} & 68.6 & 60.8 & 59.5 & 57.1 & 49.7 & 57.7 & 85.5 & 93.1 & 92.8& \textbf{69.4} \\
\includegraphics[width=0.33cm]{images/logo/huggingface-logo.png} \textbf{Gemma4-E2B} & 53.0 & 69.3 & 58.6 & 71.2 & 45.6 & 68.1 & 82.7 & 69.7 & 97.7& \textbf{68.4} \\
\includegraphics[width=0.33cm]{images/logo/huggingface-logo.png} \textbf{Gemma3-4B} & 62.1 & 58.5 & 50.0 & 50.9 & 49.2 & 51.2 & 83.2 & 90.0 & 93.2& \textbf{65.4} \\
\includegraphics[width=0.33cm]{images/logo/huggingface-logo.png} \textbf{Llama3.2-1B} & 36.2 & 59.0 & 50.9 & 46.9 & 67.6 & 86.2 & 49.2 & 81.9 & 97.0& \textbf{63.9} \\
\includegraphics[width=0.33cm]{images/logo/huggingface-logo.png} \textbf{Qwen3.5-2B} & 55.7 & 59.9 & 47.0 & 37.9 & 62.8 & 59.5 & 79.8 & 72.9 & 96.3& \textbf{63.5} \\
\includegraphics[width=0.33cm]{images/logo/huggingface-logo.png} \textbf{Gemma3-1B} & 54.2 & 54.0 & 36.1 & 33.2 & 53.5 & 66.9 & 73.5 & 84.3 & 93.8& \textbf{61.1} \\
\includegraphics[width=0.33cm]{images/logo/huggingface-logo.png} \textbf{Qwen3.5-0.8B} & 52.2 & 54.7 & 41.7 & 32.0 & 54.5 & 65.9 & 66.4 & 69.0 & 94.8& \textbf{59.0} \\
\midrule
\includegraphics[width=0.33cm]{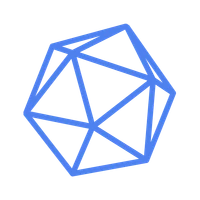} \textbf{IDA-Qwen2.5-32B} & 64.7 & 69.1 & 59.7 & 60.6 & 49.5 & 80.9 & 87.1 & 74.8 & 98.6& \textbf{71.7} \\
\includegraphics[width=0.33cm]{images/logo/cogito-logo.png} \textbf{IDA-Qwen2.5-14B} & 62.1 & 68.5 & 59.3 & 60.9 & 49.1 & 79.1 & 86.6 & 78.7 & 98.4& \textbf{71.4} \\
\includegraphics[width=0.33cm]{images/logo/deepseek-logo.png} \textbf{R1-Qwen3-8B} & 65.6 & 54.4 & 61.1 & 54.3 & 59.4 & 73.4 & 86.4 & 80.5 & 88.8& \textbf{69.3} \\
\includegraphics[width=0.33cm]{images/logo/cogito-logo.png} \textbf{IDA-Llama3.1-8B} & 38.7 & 64.1 & 54.1 & 56.3 & 55.7 & 77.5 & 89.4 & 91.5 & 93.1& \textbf{68.9} \\
\includegraphics[width=0.33cm]{images/logo/cogito-logo.png} \textbf{IDA-Llama3.2-3B} & 28.9 & 62.0 & 42.7 & 51.3 & 49.2 & 81.8 & 79.9 & 91.4 & 89.3& \textbf{64.1} \\
\includegraphics[width=0.33cm]{images/logo/deepseek-logo.png} \textbf{R1-Qwen-14B} & 7.9 & 67.7 & 64.0 & 52.4 & 46.3 & 74.4 & 30.5 & 79.4 & 83.0& \textbf{56.2} \\
\includegraphics[width=0.33cm]{images/logo/deepseek-logo.png} \textbf{R1-Llama3.1-8B} & 2.6 & 60.5 & 59.3 & 45.8 & 43.9 & 73.5 & 10.0 & 89.2 & 73.4& \textbf{50.9} \\
        \bottomrule
    \end{tabular}
    }
    \label{tab:full_leaderboard}
\end{table*}

\begin{table}[t]
    \centering
    \caption{\system Hallucination Leaderboard}
    \renewcommand{\arraystretch}{1.3} 
    \resizebox{\linewidth}{!}{
    \begin{tabular}{l|c|c|c|c|c|c|c|c}
        \toprule
        \textbf{Model} & \rotatebox{30}{\textbf{SimpleQA}} & \rotatebox{30}{\textbf{TriviaQA}} & \rotatebox{30}{\textbf{TruthfulQA}} & \rotatebox{30}{\textbf{SelfCheckGPT}} & \rotatebox{30}{\textbf{HaluEval}} & \rotatebox{30}{\textbf{FaithEval}} & \rotatebox{30}{\textbf{Vectara}} & \shortstack{\textbf{Overall} \\ \textbf{Score}} \\
        \midrule
\includegraphics[width=0.33cm]{images/logo/openai-logo.png} \textbf{GPT-5} & 72.55 & 95.43 & 83.80 & 99.58 & 74.57 & 64.34 & 97.11 & \textbf{83.91} \\
\includegraphics[width=0.33cm]{images/logo/gemini-logo-light.png} \textbf{Gemini 3} & 58.60 & 95.70 & 92.65 & 85.36 & 80.23 & 77.49 & 88.57 & \textbf{82.66} \\
\includegraphics[width=0.33cm]{images/logo/claude-logo-light.png} \textbf{Claude-4.5} & 48.08 & 92.17 & 88.24 & 97.91 & 70.97 & 72.08 & 96.03 & \textbf{80.78} \\
\includegraphics[width=0.33cm]{images/logo/openai-logo.png} \textbf{GPT-4o} & 36.92 & 95.31 & 85.91 & 89.12 & 70.07 & 67.75 & 97.59 & \textbf{77.52} \\
\includegraphics[width=0.33cm]{images/logo/claude-logo-light.png} \textbf{Claude-3.5} & 53.70 & 91.29 & 76.96 & 95.40 & 59.06 & 52.76 & 95.43 & \textbf{74.94} \\
\includegraphics[width=0.33cm]{images/logo/deepseek-logo.png} \textbf{DeepSeek-R1} & 30.93 & 100.00 & 80.64 & 86.19 & 75.04 & 58.35 & 92.90 & \textbf{74.86} \\
\includegraphics[width=0.33cm]{images/logo/grok-logo-light.png} \textbf{Grok 3} & 41.01 & 93.45 & 81.62 & 72.38 & 72.98 & 59.09 & 97.35 & \textbf{73.98} \\
\includegraphics[width=0.33cm]{images/logo/grok-logo-light.png} \textbf{Grok 4.1} & 17.64 & 92.74 & 90.81 & 13.81 & 78.32 & 77.81 & 91.34 & \textbf{66.07} \\
\includegraphics[width=0.33cm]{images/logo/grok-logo-light.png} \textbf{Grok 4} & 22.42 & 92.22 & 88.36 & 14.64 & 77.25 & 73.80 & 93.02 & \textbf{65.96} \\
        \midrule
\includegraphics[width=0.33cm]{images/logo/huggingface-logo.png} \textbf{Gemma4-31B} & 14.36 & 82.24 & 84.56 & 97.91 & 75.34 & 80.22 & 95.19 & \textbf{75.69} \\
\includegraphics[width=0.33cm]{images/logo/huggingface-logo.png} \textbf{Gemma4-26B-A4B} & 7.95 & 79.46 & 81.74 & 97.49 & 73.28 & 74.14 & 98.07 & \textbf{73.16} \\
\includegraphics[width=0.33cm]{images/logo/huggingface-logo.png} \textbf{Qwen3.5-122B-A10B} & 44.91 & 92.88 & 56.62 & 98.74 & 49.74 & 83.28 & 85.20 & \textbf{73.05} \\
\includegraphics[width=0.33cm]{images/logo/huggingface-logo.png} \textbf{Llama3.3-70B} & 36.36 & 93.30 & 75.37 & 74.90 & 68.60 & 58.03 & 97.83 & \textbf{72.06} \\
\includegraphics[width=0.33cm]{images/logo/huggingface-logo.png} \textbf{Qwen3.5-27B} & 36.13 & 88.65 & 60.42 & 99.16 & 50.24 & 83.42 & 82.19 & \textbf{71.46} \\
\includegraphics[width=0.33cm]{images/logo/huggingface-logo.png} \textbf{Qwen3.5-35B-A3B} & 38.90 & 89.12 & 51.10 & 98.74 & 49.93 & 85.83 & 80.87 & \textbf{70.64} \\
\includegraphics[width=0.33cm]{images/logo/huggingface-logo.png} \textbf{Qwen3.5-9B} & 33.66 & 84.95 & 51.35 & 99.16 & 50.33 & 86.03 & 87.85 & \textbf{70.47} \\
\includegraphics[width=0.33cm]{images/logo/huggingface-logo.png} \textbf{Qwen3.5-4B} & 33.10 & 81.59 & 48.53 & 100.00 & 50.21 & 87.09 & 80.87 & \textbf{68.77} \\
\includegraphics[width=0.33cm]{images/logo/huggingface-logo.png} \textbf{GPT-120B} & 13.82 & 85.39 & 81.50 & 43.93 & 73.38 & 58.79 & 92.30 & \textbf{64.16} \\
\includegraphics[width=0.33cm]{images/logo/huggingface-logo.png} \textbf{GPT-20B} & 18.10 & 77.40 & 74.63 & 40.34 & 66.47 & 52.74 & 88.09 & \textbf{59.68} \\
\includegraphics[width=0.33cm]{images/logo/huggingface-logo.png} \textbf{Phi-4} & 17.08 & 80.94 & 76.23 & 14.23 & 65.61 & 62.14 & 92.90 & \textbf{58.45} \\

        \bottomrule
    \end{tabular}
    }
    \label{tab:hallucination_results}
\end{table}

\begin{table}
    \centering
    \caption{\system Code Security Leaderboard}
    \renewcommand{\arraystretch}{1.3} 
    \resizebox{\linewidth}{!}{
    \begin{tabular}{l|ccccc|cccc|c}
        \toprule
        \multirow{2}{*}{\textbf{Model}} & \multicolumn{5}{c|}{\textbf{CyberSecEval 3}} & \multicolumn{4}{c|}{\textbf{SecCodePLT}} & \multirow{2}{*}{\shortstack{\textbf{Overall} \\ \textbf{Score}}} \\
        \cmidrule(lr){2-6} \cmidrule(lr){7-10}
        & \textbf{C} & \textbf{C++} & \textbf{Java} & \textbf{Php} & \textbf{Python} & \textbf{Inst} & \textbf{Auto} & \textbf{Norm} & \textbf{Aug}
\\
        \midrule
\includegraphics[width=0.33cm]{images/logo/openai-logo.png} \textbf{GPT-5} & 66.30 & 83.78 & 84.50 & 88.89 & 70.80 & 82.41 & 81.18 & 72.05 & 91.54 & \textbf{79.72} \\
\includegraphics[width=0.33cm]{images/logo/deepseek-logo.png} \textbf{DeepSeek-R1} & 91.63 & 94.79 & 85.59 & 88.58 & 92.17 & 63.59 & 60.08 & 53.99 & 69.68 & \textbf{77.11} \\
\includegraphics[width=0.33cm]{images/logo/gemini-logo-light.png} \textbf{Gemini 3} & 64.02 & 77.22 & 84.93 & 87.65 & 70.37 & 72.72 & 72.24 & 61.12 & 83.84 & \textbf{74.27} \\
\includegraphics[width=0.33cm]{images/logo/openai-logo.png} \textbf{GPT-4o} & 92.73 & 94.21 & 86.03 & 90.43 & 91.74 & 54.28 & 52.76 & 44.30 & 62.74 & \textbf{73.07} \\
\includegraphics[width=0.33cm]{images/logo/claude-logo-light.png} \textbf{Claude-3.5} & 63.22 & 82.43 & 84.93 & 90.43 & 70.66 & 66.63 & 66.54 & 57.22 & 75.95 & \textbf{72.17} \\
\includegraphics[width=0.33cm]{images/logo/grok-logo-light.png} \textbf{Grok 4.1} & 66.30 & 83.40 & 85.15 & 86.73 & 70.66 & 66.25 & 65.59 & 53.04 & 78.80 & \textbf{71.97} \\
\includegraphics[width=0.33cm]{images/logo/claude-logo-light.png} \textbf{Claude-4.5} & 56.61 & 79.34 & 84.93 & 87.96 & 70.94 & 72.81 & 55.51 & 54.66 & 73.67 & \textbf{70.26} \\
\includegraphics[width=0.33cm]{images/logo/grok-logo-light.png} \textbf{Grok 4} & 62.78 & 80.12 & 84.93 & 89.20 & 69.80 & 64.26 & 61.98 & 48.38 & 77.85 & \textbf{69.73} \\
\includegraphics[width=0.33cm]{images/logo/grok-logo-light.png} \textbf{Grok 3} & 60.13 & 83.40 & 85.15 & 91.36 & 71.08 & 62.36 & 55.99 & 44.30 & 74.05 & \textbf{68.60} \\
        \midrule
\includegraphics[width=0.33cm]{images/logo/huggingface-logo.png} \textbf{GPT-120B} & 68.72 & 84.75 & 85.37 & 89.20 & 72.65 & 80.70 & 80.23 & 69.87 & 91.06 & \textbf{80.10} \\
\includegraphics[width=0.33cm]{images/logo/huggingface-logo.png} \textbf{GPT-20B} & 71.37 & 88.03 & 85.81 & 91.36 & 77.21 & 73.48 & 73.95 & 62.74 & 84.70 & \textbf{78.18} \\
\includegraphics[width=0.33cm]{images/logo/huggingface-logo.png} \textbf{Gemma4-26B-A4B} & 63.66 & 82.24 & 86.68 & 93.21 & 74.36 & 73.86 & 73.86 & 61.98 & 85.74 & \textbf{76.42} \\
\includegraphics[width=0.33cm]{images/logo/huggingface-logo.png} \textbf{Gemma4-31B} & 66.08 & 81.47 & 86.03 & 91.98 & 73.50 & 71.58 & 71.20 & 59.51 & 83.27 & \textbf{75.27} \\
\includegraphics[width=0.33cm]{images/logo/huggingface-logo.png} \textbf{Qwen3.5-27B} & 84.80 & 91.51 & 90.17 & 92.28 & 84.47 & 67.87 & 54.09 & 52.85 & 69.11 & \textbf{74.95} \\
\includegraphics[width=0.33cm]{images/logo/huggingface-logo.png} \textbf{Qwen3.5-122B-A10B} & 84.36 & 92.28 & 90.39 & 93.21 & 84.05 & 65.97 & 51.71 & 52.19 & 65.49 & \textbf{74.14} \\
\includegraphics[width=0.33cm]{images/logo/huggingface-logo.png} \textbf{Qwen3.5-9B} & 88.55 & 95.37 & 93.89 & 95.68 & 87.18 & 60.84 & 47.15 & 46.67 & 61.31 & \textbf{73.15} \\
\includegraphics[width=0.33cm]{images/logo/huggingface-logo.png} \textbf{Qwen3.5-35B-A3B} & 87.22 & 93.82 & 90.17 & 92.59 & 86.18 & 59.60 & 47.34 & 47.53 & 59.41 & \textbf{72.05} \\
\includegraphics[width=0.33cm]{images/logo/huggingface-logo.png} \textbf{Qwen3.5-4B} & 90.09 & 94.59 & 89.74 & 95.99 & 86.18 & 57.03 & 45.63 & 42.11 & 60.55 & \textbf{71.53} \\
\includegraphics[width=0.33cm]{images/logo/huggingface-logo.png} \textbf{Llama3.3-70B} & 57.93 & 81.47 & 85.59 & 91.67 & 69.66 & 60.17 & 59.41 & 45.25 & 74.33 & \textbf{68.41} \\
\includegraphics[width=0.33cm]{images/logo/huggingface-logo.png} \textbf{Phi-4} & 62.33 & 80.89 & 84.93 & 90.74 & 73.08 & 57.22 & 57.13 & 44.11 & 70.25 & \textbf{67.99} \\

        \bottomrule
    \end{tabular}
    }
    \label{tab:code_security_results}
\end{table}

\subsection{Hallucination}
\label{res:hallucination}

Table~\ref{tab:hallucination_results} reveals GPT-5 achieving the highest overall hallucination resilience. Gemini~3 and Claude-4.5 follow closely, reflecting steady improvements in factual reliability and internal consistency compared to earlier generations. The high \textit{SelfCheckGPT} scores across top systems indicate strong self-verification capabilities, though the disparity between these and factuality-oriented tests like \textit{SimpleQA} and \textit{TruthfulQA} underscores that self-consistency alone does not ensure truthfulness. Among open-weight models, Gemma4 and Qwen3.5 variants are competitive on knowledge-dense evaluations such as \textit{FaithEval}, while GPT-120B and Phi-4 show steep degradation in \textit{SelfCheckGPT}. The notably high scores of DeepSeek-R1 demonstrate the growing strength of safety-tuned open-weight models.

\subsection{Code Security}\label{res:code}

Table~\ref{tab:code_security_results} shows GPT-120B and GPT-20B leading among open-weight models, while GPT-5 and DeepSeek-R1 top the API group. Most models score well on \textit{CyberSecEval}, indicating broad competence in generating vulnerability-free code, but the more demanding \textit{SecCodePLT} test exposes substantial degradation, especially in the setting where models must correctly complete partially insecure or ambiguous templates. Security-policy augmentation produces moderate gains for all models, suggesting successful policy internalization. Notably, Qwen3.5 models exhibit a reversed pattern: strong on \textit{CyberSecEval} but weaker on \textit{SecCodePLT}, suggesting a different calibration tradeoff.
\subsection{Safety Alignment}\label{res:alignment}

Table~\ref{tab:safety_alignment_results} shows uniformly strong performance among proprietary models, with consistently high refusal across tests indicating well-calibrated safety alignment. Across open-weight models, the strongest performers are Phi-4, GPT-120B, and the Gemma4 variants, matching or exceeding the performance of proprietary models. Notably, Qwen3.5 models achieve near-perfect scores on several guardrail tests but show a marked gap on \textit{OpenAI Moderation} and \textit{Perspective API}, suggesting uneven coverage across safety taxonomies. Grok~4.1 stands out among API models for its lower alignment scores, indicating possible under-tuning on certain safety dimensions.

\begin{table}
    \centering
    \caption{\system Safety Alignment Leaderboard}
    \renewcommand{\arraystretch}{1.3}
    \begin{adjustbox}{max width=\columnwidth}
    \begin{tabular}{l|c|c|c|c|c|c|c|c|c|c}
        \toprule
        \textbf{Model}
        & \rotatebox{60}{\shortstack{\textbf{Llama} \\ \textbf{Guard 1}}}
        & \rotatebox{60}{\shortstack{\textbf{Llama} \\ \textbf{Guard 2}}}
        & \rotatebox{60}{\shortstack{\textbf{Llama} \\ \textbf{Guard 3}}}
        & \rotatebox{60}{\shortstack{\textbf{Llama} \\ \textbf{Guard 4}}}
        & \rotatebox{60}{\shortstack{\textbf{OpenAI} \\ \textbf{Moderation}}}
        & \rotatebox{60}{\shortstack{\textbf{Perspective} \\ \textbf{API}}}
        & \rotatebox{60}{\shortstack{\textbf{Wild} \\ \textbf{Guard}}}
        & \rotatebox{60}{\shortstack{\textbf{Simple} \\ \textbf{Safety}}}
        & \rotatebox{60}{\shortstack{\textbf{Anthropic} \\ \textbf{Redteam}}}
        & \shortstack{\textbf{Overall} \\ \textbf{Score}} \\
        \midrule
\includegraphics[width=0.33cm]{images/logo/claude-logo-light.png} \textbf{Claude-3.5} & 100.00 & 99.61 & 100.00 & 100.00 & 97.13 & 98.20 & 99.34 & 100.00 & 99.10 & \textbf{99.26} \\
\includegraphics[width=0.33cm]{images/logo/openai-logo.png} \textbf{GPT-4o} & 99.40 & 98.18 & 98.55 & 97.82 & 99.04 & 96.62 & 89.79 & 100.00 & 98.00 & \textbf{97.49} \\
\includegraphics[width=0.33cm]{images/logo/deepseek-logo.png} \textbf{DeepSeek-R1} & 99.40 & 96.74 & 98.73 & 99.09 & 98.47 & 95.24 & 92.71 & 100.00 & 96.50 & \textbf{97.43} \\
\includegraphics[width=0.33cm]{images/logo/claude-logo-light.png} \textbf{Claude-4.5} & 99.20 & 97.53 & 100.00 & 99.45 & 92.15 & 94.93 & 99.73 & 100.00 & 90.40 & \textbf{97.04} \\
\includegraphics[width=0.33cm]{images/logo/grok-logo-light.png} \textbf{Grok 4} & 99.80 & 98.05 & 96.75 & 99.45 & 91.95 & 90.49 & 96.82 & 100.00 & 99.70 & \textbf{97.00} \\
\includegraphics[width=0.33cm]{images/logo/openai-logo.png} \textbf{GPT-5} & 100.00 & 93.58 & 99.27 & 100.00 & 87.64 & 86.23 & 96.46 & 100.00 & 97.58 & \textbf{95.64} \\
\includegraphics[width=0.33cm]{images/logo/grok-logo-light.png} \textbf{Grok 3} & 99.60 & 97.01 & 85.71 & 99.82 & 95.21 & 87.10 & 82.49 & 100.00 & 98.90 & \textbf{93.98} \\
\includegraphics[width=0.33cm]{images/logo/gemini-logo-light.png} \textbf{Gemini 3} & 99.00 & 91.67 & 94.21 & 95.45 & 90.80 & 87.74 & 90.45 & 100.00 & 89.60 & \textbf{93.21} \\
\includegraphics[width=0.33cm]{images/logo/grok-logo-light.png} \textbf{Grok 4.1} & 97.60 & 93.10 & 98.37 & 98.18 & 63.46 & 66.35 & 86.53 & 100.00 & 88.35 & \textbf{87.99} \\
        \midrule
\includegraphics[width=0.33cm]{images/logo/huggingface-logo.png} \textbf{Phi-4} & 100.00 & 98.57 & 99.46 & 99.45 & 98.85 & 98.41 & 97.88 & 100.00 & 98.10 & \textbf{98.97} \\
\includegraphics[width=0.33cm]{images/logo/huggingface-logo.png} \textbf{GPT-120B} & 99.80 & 96.09 & 100.00 & 100.00 & 97.51 & 92.18 & 99.34 & 100.00 & 98.90 & \textbf{98.20} \\
\includegraphics[width=0.33cm]{images/logo/huggingface-logo.png} \textbf{GPT-20B} & 99.80 & 94.92 & 100.00 & 99.82 & 96.17 & 90.77 & 98.01 & 100.00 & 97.90 & \textbf{97.49} \\
\includegraphics[width=0.33cm]{images/logo/huggingface-logo.png} \textbf{Gemma4-31B} & 100.00 & 95.96 & 100.00 & 100.00 & 94.44 & 92.49 & 97.61 & 100.00 & 96.50 & \textbf{97.45} \\
\includegraphics[width=0.33cm]{images/logo/huggingface-logo.png} \textbf{Gemma4-26B-A4B} & 100.00 & 95.44 & 99.82 & 100.00 & 93.68 & 92.60 & 98.28 & 100.00 & 96.30 & \textbf{97.35} \\
\includegraphics[width=0.33cm]{images/logo/huggingface-logo.png} \textbf{Llama3.3-70B} & 98.80 & 96.88 & 92.77 & 97.64 & 88.51 & 92.92 & 86.47 & 100.00 & 92.50 & \textbf{94.05} \\
\includegraphics[width=0.33cm]{images/logo/huggingface-logo.png} \textbf{Qwen3.5-9B} & 99.60 & 98.83 & 99.82 & 100.00 & 47.51 & 34.25 & 100.00 & 100.00 & 99.60 & \textbf{86.62} \\
\includegraphics[width=0.33cm]{images/logo/huggingface-logo.png} \textbf{Qwen3.5-27B} & 99.20 & 99.35 & 99.64 & 99.82 & 42.15 & 31.40 & 100.00 & 100.00 & 98.80 & \textbf{85.59} \\
\includegraphics[width=0.33cm]{images/logo/huggingface-logo.png} \textbf{Qwen3.5-35B-A3B} & 99.20 & 98.57 & 99.28 & 100.00 & 42.53 & 30.55 & 100.00 & 100.00 & 99.50 & \textbf{85.51} \\
\includegraphics[width=0.33cm]{images/logo/huggingface-logo.png} \textbf{Qwen3.5-122B-A10B} & 98.80 & 98.44 & 99.46 & 100.00 & 43.10 & 29.49 & 100.00 & 100.00 & 99.50 & \textbf{85.42} \\
\includegraphics[width=0.33cm]{images/logo/huggingface-logo.png} \textbf{Qwen3.5-4B} & 99.80 & 98.18 & 99.10 & 100.00 & 41.00 & 29.49 & 100.00 & 100.00 & 99.20 & \textbf{85.20} \\

        \bottomrule
    \end{tabular}
    \end{adjustbox}
    \label{tab:safety_alignment_results}
\end{table}

\subsection{Over Refusal}\label{res:refusal}

Table~\ref{tab:over_refusal_results} shows strong responsiveness to innocuous queries across the board, with Gemini~3, Grok~3, and GPT-4o ranking among the top performers. These models maintain high acceptance rates across all four tests, indicating effective calibration across standard, categorical, and semantically tricky prompts. In contrast, Claude-3.5 displays substantially lower \textit{OR Bench} scores, suggesting a tendency toward over-cautious refusals in contexts that resemble but do not constitute unsafe content. Among open-weight models, Llama3.3-70B leads, while GPT-120B and GPT-20B show notably low \textit{OR Bench} scores despite strong performance on other tests.

\begin{table}
    \centering
    \caption{\system Over Refusal Leaderboard}
    \renewcommand{\arraystretch}{1.3}
    \begin{adjustbox}{max width=\columnwidth}
    \begin{tabular}{l|c|c|c|c|c}
        \toprule
        \textbf{Model} & \textbf{OK Test} & \textbf{OR Bench} & \textbf{Wild Guard} & \textbf{XS Test} & \shortstack{\textbf{Overall} \\ \textbf{Score}} \\
        \midrule
\includegraphics[width=0.33cm]{images/logo/gemini-logo-light.png} \textbf{Gemini 3} & 98.86 & 91.21 & 98.76 & 88.89 & \textbf{94.43} \\
\includegraphics[width=0.33cm]{images/logo/grok-logo-light.png} \textbf{Grok 3} & 97.71 & 97.95 & 98.04 & 84.00 & \textbf{94.43} \\
\includegraphics[width=0.33cm]{images/logo/openai-logo.png} \textbf{GPT-4o} & 97.14 & 92.57 & 97.63 & 90.22 & \textbf{94.39} \\
\includegraphics[width=0.33cm]{images/logo/deepseek-logo.png} \textbf{DeepSeek-R1} & 97.71 & 77.18 & 98.76 & 92.44 & \textbf{91.53} \\
\includegraphics[width=0.33cm]{images/logo/grok-logo-light.png} \textbf{Grok 4} & 99.14 & 71.42 & 90.42 & 92.22 & \textbf{88.30} \\
\includegraphics[width=0.33cm]{images/logo/grok-logo-light.png} \textbf{Grok 4.1} & 98.57 & 67.40 & 94.95 & 92.00 & \textbf{88.23} \\
\includegraphics[width=0.33cm]{images/logo/openai-logo.png} \textbf{GPT-5} & 96.84 & 65.88 & 95.26 & 93.50 & \textbf{87.87} \\
\includegraphics[width=0.33cm]{images/logo/claude-logo-light.png} \textbf{Claude-4.5} & 91.43 & 80.74 & 76.31 & 80.89 & \textbf{82.34} \\
\includegraphics[width=0.33cm]{images/logo/claude-logo-light.png} \textbf{Claude-3.5} & 73.43 & 20.32 & 83.83 & 91.33 & \textbf{67.23} \\
        \midrule
\includegraphics[width=0.33cm]{images/logo/huggingface-logo.png} \textbf{Llama3.3-70B} & 93.14 & 95.00 & 97.53 & 90.22 & \textbf{93.97} \\
\includegraphics[width=0.33cm]{images/logo/huggingface-logo.png} \textbf{Phi-4} & 98.00 & 62.02 & 95.78 & 89.56 & \textbf{86.34} \\
\includegraphics[width=0.33cm]{images/logo/huggingface-logo.png} \textbf{Gemma4-31B} & 93.71 & 61.03 & 94.23 & 92.22 & \textbf{85.30} \\
\includegraphics[width=0.33cm]{images/logo/huggingface-logo.png} \textbf{Gemma4-26B-A4B} & 91.71 & 52.62 & 93.61 & 91.33 & \textbf{82.32} \\
\includegraphics[width=0.33cm]{images/logo/huggingface-logo.png} \textbf{Qwen3.5-27B} & 96.86 & 43.15 & 87.11 & 91.33 & \textbf{79.61} \\
\includegraphics[width=0.33cm]{images/logo/huggingface-logo.png} \textbf{Qwen3.5-122B-A10B} & 96.26 & 38.49 & 88.72 & 92.44 & \textbf{78.98} \\
\includegraphics[width=0.33cm]{images/logo/huggingface-logo.png} \textbf{Qwen3.5-9B} & 95.42 & 41.26 & 86.67 & 91.33 & \textbf{78.67} \\
\includegraphics[width=0.33cm]{images/logo/huggingface-logo.png} \textbf{Qwen3.5-4B} & 95.99 & 36.59 & 88.52 & 92.79 & \textbf{78.47} \\
\includegraphics[width=0.33cm]{images/logo/huggingface-logo.png} \textbf{Qwen3.5-35B-A3B} & 96.55 & 37.31 & 87.40 & 92.19 & \textbf{78.36} \\
\includegraphics[width=0.33cm]{images/logo/huggingface-logo.png} \textbf{GPT-20B} & 95.42 & 23.43 & 87.31 & 91.78 & \textbf{74.48} \\
\includegraphics[width=0.33cm]{images/logo/huggingface-logo.png} \textbf{GPT-120B} & 95.71 & 14.78 & 87.02 & 92.22 & \textbf{72.44} \\

        \bottomrule
    \end{tabular}
    \end{adjustbox}
    \label{tab:over_refusal_results}
\end{table}

\subsection{Adversarial Robustness}

Table~\ref{tab:adversarial_results} reveals significant variation in LLM responses to prompt attacks and subtle perturbations. Gemini~3 leads by a wide margin, while most API models cluster between 61--72 points. Larger models performed better in clean conditions (\textit{AdvGlue}), but their advantage decreased notably on adversarially perturbed prompts (\textit{AdvGlue++}). Among open-weight models, Gemma4-31B leads with 75.2, outperforming all API models except Gemini~3. This suggests that scaling alone is insufficient and must be complemented with carefully designed training or alignment.

\begin{table}
    \centering
    \caption{\system Adversarial Robustness Leaderboard}
    \renewcommand{\arraystretch}{1.3}
    \resizebox{\linewidth}{!}{
    \begin{tabular}{l|ccccc|ccccc|c}
        \toprule
        \multirow{2}{*}{\textbf{Model}} & \multicolumn{5}{c|}{\textbf{AdvGlue}} & \multicolumn{5}{c|}{\textbf{AdvGlue++}} & \multirow{2}{*}{\shortstack{\textbf{Overall} \\ \textbf{Score}}} \\
        \cmidrule(lr){2-6} \cmidrule(lr){7-11}
        & \textbf{MNLI} & \textbf{QNLI} & \textbf{QQP} & \textbf{RTE} & \textbf{SST2}
        & \textbf{MNLI} & \textbf{QNLI} & \textbf{QQP} & \textbf{RTE} & \textbf{SST2} &  \\
        \midrule
\includegraphics[width=0.33cm]{images/logo/gemini-logo-light.png} \textbf{Gemini 3} & 86.78 & 81.76 & 83.33 & 90.12 & 85.14 & 69.68 & 74.94 & 60.81 & 52.76 & 86.22 & \textbf{77.15} \\
\includegraphics[width=0.33cm]{images/logo/openai-logo.png} \textbf{GPT-5} & 73.55 & 80.41 & 79.49 & 90.12 & 72.30 & 47.09 & 82.89 & 55.26 & 52.80 & 82.06 & \textbf{71.60} \\
\includegraphics[width=0.33cm]{images/logo/grok-logo-light.png} \textbf{Grok 4} & 65.29 & 79.73 & 80.77 & 96.30 & 79.05 & 31.97 & 77.28 & 54.63 & 54.99 & 83.41 & \textbf{70.34} \\
\includegraphics[width=0.33cm]{images/logo/grok-logo-light.png} \textbf{Grok 4.1} & 61.16 & 80.41 & 76.92 & 92.59 & 79.05 & 35.79 & 73.99 & 51.19 & 48.52 & 83.01 & \textbf{68.26} \\
\includegraphics[width=0.33cm]{images/logo/deepseek-logo.png} \textbf{DeepSeek-R1} & 74.38 & 79.05 & 75.64 & 91.36 & 63.51 & 60.91 & 59.81 & 45.52 & 43.10 & 78.54 & \textbf{67.18} \\
\includegraphics[width=0.33cm]{images/logo/grok-logo-light.png} \textbf{Grok 3} & 53.72 & 75.00 & 75.64 & 90.12 & 79.73 & 33.76 & 70.69 & 49.31 & 51.89 & 82.79 & \textbf{66.26} \\
\includegraphics[width=0.33cm]{images/logo/claude-logo-light.png} \textbf{Claude-3.5} & 52.07 & 82.43 & 82.05 & 81.48 & 62.84 & 39.00 & 63.39 & 51.32 & 41.91 & 70.23 & \textbf{62.67} \\
\includegraphics[width=0.33cm]{images/logo/openai-logo.png} \textbf{GPT-4o} & 50.41 & 77.03 & 71.79 & 90.12 & 66.22 & 35.81 & 60.99 & 48.92 & 46.70 & 77.90 & \textbf{62.59} \\
\includegraphics[width=0.33cm]{images/logo/claude-logo-light.png} \textbf{Claude-4.5} & 53.72 & 82.43 & 67.95 & 88.89 & 67.57 & 27.62 & 60.15 & 48.38 & 45.74 & 73.81 & \textbf{61.63} \\
        \midrule
\includegraphics[width=0.33cm]{images/logo/huggingface-logo.png} \textbf{Gemma4-31B} & 79.34 & 85.81 & 84.62 & 87.65 & 79.05 & 55.56 & 78.55 & 61.17 & 57.72 & 82.70 & \textbf{75.22} \\
\includegraphics[width=0.33cm]{images/logo/huggingface-logo.png} \textbf{Gemma4-26B-A4B} & 68.60 & 78.38 & 78.21 & 90.12 & 73.65 & 52.83 & 74.98 & 54.52 & 50.02 & 80.06 & \textbf{70.14} \\
\includegraphics[width=0.33cm]{images/logo/huggingface-logo.png} \textbf{Qwen3.5-27B} & 66.94 & 87.16 & 83.33 & 83.95 & 68.24 & 57.29 & 79.78 & 55.61 & 43.64 & 72.17 & \textbf{69.81} \\
\includegraphics[width=0.33cm]{images/logo/huggingface-logo.png} \textbf{Qwen3.5-122B-A10B} & 71.90 & 81.76 & 80.77 & 77.78 & 67.57 & 60.08 & 79.53 & 56.15 & 43.46 & 73.00 & \textbf{69.20} \\
\includegraphics[width=0.33cm]{images/logo/huggingface-logo.png} \textbf{Qwen3.5-35B-A3B} & 63.64 & 84.46 & 83.33 & 80.25 & 60.14 & 62.75 & 79.96 & 55.41 & 42.19 & 68.59 & \textbf{68.07} \\
\includegraphics[width=0.33cm]{images/logo/huggingface-logo.png} \textbf{Llama3.3-70B} & 61.98 & 79.05 & 79.49 & 86.42 & 80.41 & 29.54 & 71.13 & 54.74 & 51.80 & 80.97 & \textbf{67.55} \\
\includegraphics[width=0.33cm]{images/logo/huggingface-logo.png} \textbf{GPT-120B} & 57.85 & 85.14 & 87.18 & 85.19 & 66.89 & 30.46 & 72.84 & 56.36 & 47.47 & 74.08 & \textbf{66.34} \\
\includegraphics[width=0.33cm]{images/logo/huggingface-logo.png} \textbf{Qwen3.5-9B} & 61.98 & 81.76 & 76.92 & 80.25 & 56.76 & 58.02 & 79.41 & 54.09 & 44.01 & 67.83 & \textbf{66.10} \\
\includegraphics[width=0.33cm]{images/logo/huggingface-logo.png} \textbf{GPT-20B} & 64.46 & 81.08 & 80.77 & 86.42 & 61.49 & 41.03 & 73.30 & 54.67 & 45.97 & 70.17 & \textbf{65.94} \\
\includegraphics[width=0.33cm]{images/logo/huggingface-logo.png} \textbf{Qwen3.5-4B} & 51.24 & 80.41 & 78.21 & 83.95 & 56.08 & 60.24 & 77.19 & 54.83 & 42.23 & 66.37 & \textbf{65.07} \\
\includegraphics[width=0.33cm]{images/logo/huggingface-logo.png} \textbf{Phi-4} & 59.50 & 72.97 & 78.21 & 87.65 & 62.84 & 40.84 & 62.09 & 50.63 & 45.97 & 73.83 & \textbf{63.45} \\

        \bottomrule
    \end{tabular}
    }
    \label{tab:adversarial_results}
\end{table}

\subsection{Jailbreak Robustness}

Table~\ref{tab:jailbreak_results} reveals stark differences in models’ ability to withstand jailbreak attempts. Claude-4.5 leads by a wide margin, demonstrating exceptional resilience across all attack types. GPT-5 ranks second, performing strongly on \textit{Jailbroken} and \textit{Cipher}, but showing a moderate decline under the adaptive PAIR attack. Most other proprietary models maintain robustness to direct jailbreaks but perform poorly on \textit{PAIR}, underscoring that adaptive attacks remain a major vulnerability. Remarkably, Qwen3.5 models achieve near-perfect jailbreak robustness (98--99), outperforming all API models including Claude-4.5.

\begin{table}
    \centering
    \caption{\system Jailbreak Robustness Leaderboard}
    % \scriptsize
    \begin{tabular}{l|c|c|c|c}
        \toprule
        \textbf{Model}
        & \textbf{PAIR}
        & \textbf{Jailbroken}
        & \textbf{Cipher}
        & \shortstack{\textbf{Overall} \\ \textbf{Score}} \\
        \midrule
\includegraphics[width=0.33cm]{images/logo/claude-logo-light.png} \textbf{Claude-4.5} & 96.00 & 99.52 & 99.88 & \textbf{98.46} \\
\includegraphics[width=0.33cm]{images/logo/openai-logo.png} \textbf{GPT-5} & 75.00 & 98.82 & 96.80 & \textbf{90.21} \\
\includegraphics[width=0.33cm]{images/logo/grok-logo-light.png} \textbf{Grok 4.1} & 38.00 & 99.10 & 97.12 & \textbf{78.07} \\
\includegraphics[width=0.33cm]{images/logo/grok-logo-light.png} \textbf{Grok 4} & 23.00 & 96.77 & 99.72 & \textbf{73.17} \\
\includegraphics[width=0.33cm]{images/logo/claude-logo-light.png} \textbf{Claude-3.5} & 41.00 & 99.07 & 71.55 & \textbf{70.54} \\
\includegraphics[width=0.33cm]{images/logo/gemini-logo-light.png} \textbf{Gemini 3} & 20.00 & 94.90 & 96.15 & \textbf{70.35} \\
\includegraphics[width=0.33cm]{images/logo/deepseek-logo.png} \textbf{DeepSeek-R1} & 11.11 & 83.13 & 61.13 & \textbf{51.79} \\
\includegraphics[width=0.33cm]{images/logo/openai-logo.png} \textbf{GPT-4o} & 5.00 & 84.21 & 64.85 & \textbf{51.35} \\
\includegraphics[width=0.33cm]{images/logo/grok-logo-light.png} \textbf{Grok 3} & 0.00 & 68.41 & 73.67 & \textbf{47.36} \\
        \midrule
\includegraphics[width=0.33cm]{images/logo/huggingface-logo.png} \textbf{Qwen3.5-27B} & 99.00 & 99.72 & 99.65 & \textbf{99.46} \\
\includegraphics[width=0.33cm]{images/logo/huggingface-logo.png} \textbf{Qwen3.5-122B-A10B} & 98.00 & 99.83 & 99.90 & \textbf{99.24} \\
\includegraphics[width=0.33cm]{images/logo/huggingface-logo.png} \textbf{Qwen3.5-9B} & 98.00 & 99.79 & 98.72 & \textbf{98.84} \\
\includegraphics[width=0.33cm]{images/logo/huggingface-logo.png} \textbf{Qwen3.5-35B-A3B} & 98.00 & 99.93 & 98.30 & \textbf{98.74} \\
\includegraphics[width=0.33cm]{images/logo/huggingface-logo.png} \textbf{Qwen3.5-4B} & 97.00 & 99.66 & 99.05 & \textbf{98.57} \\
\includegraphics[width=0.33cm]{images/logo/huggingface-logo.png} \textbf{GPT-120B} & 88.00 & 99.79 & 87.98 & \textbf{91.92} \\
\includegraphics[width=0.33cm]{images/logo/huggingface-logo.png} \textbf{GPT-20B} & 72.00 & 99.64 & 91.37 & \textbf{87.67} \\
\includegraphics[width=0.33cm]{images/logo/huggingface-logo.png} \textbf{Gemma4-31B} & 47.00 & 98.34 & 75.42 & \textbf{73.59} \\
\includegraphics[width=0.33cm]{images/logo/huggingface-logo.png} \textbf{Gemma4-26B-A4B} & 48.00 & 98.55 & 71.60 & \textbf{72.72} \\
\includegraphics[width=0.33cm]{images/logo/huggingface-logo.png} \textbf{Phi-4} & 10.00 & 96.38 & 63.92 & \textbf{56.77} \\
\includegraphics[width=0.33cm]{images/logo/huggingface-logo.png} \textbf{Llama3.3-70B} & 2.00 & 78.72 & 62.70 & \textbf{47.81} \\

        \bottomrule
    \end{tabular}
    \label{tab:jailbreak_results}
\end{table}

\subsection{OOD Robustness}
\label{res:ood}

Table~\ref{tab:ood_robustness_results} shows that most models exhibit consistently high accuracy across both stylistic rewrites and perturbation intensities, demonstrating robust semantic stability even under substantial surface-level drift. Gemini~3 leads overall, followed closely by the Grok models. Among open-weight systems, Gemma4 and Llama3.3-70B match or exceed several API models. Across all model families, performance is generally stronger on word-level augmentations than on style transfers, particularly under higher perturbation, where stylistic transformations impose deeper syntactic and lexical shifts.

\begin{table}
    \centering
    \caption{\system OOD Robustness Leaderboard}
    \renewcommand{\arraystretch}{1.3} 
    \resizebox{\linewidth}{!}{
    \begin{tabular}{l|cc|cc|cc|cc|cc|c}
        \toprule
        \multirow{2}{*}{\textbf{Model}} & \multicolumn{2}{c|}{\textbf{Word Level}} & \multicolumn{2}{c|}{\textbf{Bible}} & \multicolumn{2}{c|}{\textbf{Romantic}} & \multicolumn{2}{c|}{\textbf{Shakespeare}} & \multicolumn{2}{c|}{\textbf{Tweet}} & \multirow{2}{*}{\shortstack{\textbf{Overall} \\ \textbf{Score}}}\\
        \cmidrule(lr){2-3} \cmidrule(lr){4-5} \cmidrule(lr){6-7} \cmidrule(lr){8-9} \cmidrule(lr){10-11}
        & \textbf{Aug} & \textbf{ShaW} & \textbf{p=0} & \textbf{p=0.6} & \textbf{p=0} & \textbf{p=0.6} & \textbf{p=0} & \textbf{p=0.6} & \textbf{p=0} & \textbf{p=0.6} \\
        \midrule
\includegraphics[width=0.33cm]{images/logo/gemini-logo-light.png} \textbf{Gemini 3} & 97.02 & 95.64 & 89.11 & 86.58 & 89.91 & 88.99 & 92.66 & 88.42 & 93.92 & 93.69 & \textbf{91.59} \\
\includegraphics[width=0.33cm]{images/logo/grok-logo-light.png} \textbf{Grok 4} & 95.30 & 93.58 & 87.84 & 84.40 & 88.30 & 87.39 & 91.74 & 86.93 & 91.86 & 92.32 & \textbf{89.97} \\
\includegraphics[width=0.33cm]{images/logo/grok-logo-light.png} \textbf{Grok 4.1} & 94.95 & 93.58 & 87.96 & 84.29 & 87.84 & 87.39 & 91.17 & 86.70 & 92.09 & 91.74 & \textbf{89.77} \\
\includegraphics[width=0.33cm]{images/logo/grok-logo-light.png} \textbf{Grok 3} & 94.61 & 92.89 & 87.50 & 84.29 & 87.73 & 87.96 & 90.60 & 85.78 & 92.66 & 91.86 & \textbf{89.59} \\
\includegraphics[width=0.33cm]{images/logo/openai-logo.png} \textbf{GPT-5} & 94.50 & 92.32 & 87.84 & 82.91 & 86.93 & 87.39 & 91.86 & 87.16 & 93.12 & 90.60 & \textbf{89.46} \\
\includegraphics[width=0.33cm]{images/logo/claude-logo-light.png} \textbf{Claude-4.5} & 93.92 & 91.40 & 86.58 & 82.68 & 85.09 & 86.70 & 90.25 & 83.83 & 90.60 & 89.56 & \textbf{88.06} \\
\includegraphics[width=0.33cm]{images/logo/claude-logo-light.png} \textbf{Claude-3.5} & 91.63 & 90.60 & 85.44 & 81.54 & 83.26 & 85.44 & 88.76 & 84.29 & 89.68 & 89.33 & \textbf{87.00} \\
\includegraphics[width=0.33cm]{images/logo/openai-logo.png} \textbf{GPT-4o} & 92.89 & 90.14 & 84.29 & 81.31 & 84.52 & 85.78 & 88.30 & 82.34 & 90.37 & 89.56 & \textbf{86.95} \\
\includegraphics[width=0.33cm]{images/logo/deepseek-logo.png} \textbf{DeepSeek-R1} & 91.97 & 90.25 & 85.44 & 80.05 & 82.80 & 83.60 & 88.53 & 82.00 & 88.99 & 88.65 & \textbf{86.23} \\
        \midrule
\includegraphics[width=0.33cm]{images/logo/huggingface-logo.png} \textbf{Gemma4-26B-A4B} & 94.15 & 92.78 & 87.27 & 84.98 & 86.70 & 88.42 & 90.94 & 85.67 & 92.66 & 91.51 & \textbf{89.51} \\
\includegraphics[width=0.33cm]{images/logo/huggingface-logo.png} \textbf{Gemma4-31B} & 94.27 & 92.20 & 87.39 & 83.72 & 87.04 & 87.04 & 90.60 & 85.89 & 91.74 & 90.71 & \textbf{89.06} \\
\includegraphics[width=0.33cm]{images/logo/huggingface-logo.png} \textbf{Llama3.3-70B} & 93.58 & 91.51 & 86.93 & 82.80 & 86.12 & 86.24 & 90.25 & 84.63 & 92.32 & 91.28 & \textbf{88.57} \\
\includegraphics[width=0.33cm]{images/logo/huggingface-logo.png} \textbf{Phi-4} & 91.40 & 91.51 & 85.21 & 81.31 & 84.86 & 86.12 & 89.33 & 84.06 & 89.22 & 89.56 & \textbf{87.26} \\
\includegraphics[width=0.33cm]{images/logo/huggingface-logo.png} \textbf{Qwen3.5-122B-A10B} & 93.58 & 90.71 & 84.29 & 79.70 & 83.72 & 83.60 & 89.56 & 83.37 & 91.17 & 89.45 & \textbf{86.92} \\
\includegraphics[width=0.33cm]{images/logo/huggingface-logo.png} \textbf{Qwen3.5-27B} & 93.23 & 90.25 & 84.75 & 80.16 & 83.60 & 83.49 & 88.19 & 81.42 & 90.37 & 89.45 & \textbf{86.49} \\
\includegraphics[width=0.33cm]{images/logo/huggingface-logo.png} \textbf{GPT-120B} & 90.60 & 88.07 & 82.45 & 80.73 & 83.72 & 80.28 & 86.12 & 81.77 & 88.07 & 88.30 & \textbf{85.01} \\
\includegraphics[width=0.33cm]{images/logo/huggingface-logo.png} \textbf{Qwen3.5-35B-A3B} & 92.78 & 88.30 & 82.22 & 78.44 & 81.65 & 82.34 & 87.27 & 78.90 & 90.14 & 87.39 & \textbf{84.94} \\
\includegraphics[width=0.33cm]{images/logo/huggingface-logo.png} \textbf{Qwen3.5-9B} & 91.40 & 86.70 & 81.65 & 76.03 & 81.31 & 81.31 & 87.04 & 80.50 & 88.07 & 87.27 & \textbf{84.13} \\
\includegraphics[width=0.33cm]{images/logo/huggingface-logo.png} \textbf{GPT-20B} & 88.07 & 86.12 & 82.45 & 79.36 & 81.77 & 80.28 & 85.67 & 79.13 & 87.73 & 86.35 & \textbf{83.69} \\
\includegraphics[width=0.33cm]{images/logo/huggingface-logo.png} \textbf{Qwen3.5-4B} & 87.50 & 86.35 & 79.36 & 76.72 & 79.01 & 79.13 & 84.86 & 77.06 & 87.73 & 86.01 & \textbf{82.37} \\

        \bottomrule
    \end{tabular}
    }
    \label{tab:ood_robustness_results}
\end{table}

\begin{table}
    \centering
    \caption{\system Model \& Data Privacy Leaderboard}
    \renewcommand{\arraystretch}{1.3} 
    \begin{adjustbox}{max width=\columnwidth}
    \begin{tabular}{l|cc|c|cc|ccc|c}
        \toprule
        \multirow{2}{*}{\shortstack{\textbf{Model}}}  
        & \multicolumn{2}{c|}{\textbf{PII Awareness}} 
        & \multirow{2}{*}{\shortstack{\textbf{ConfAIde}}}
        & \multicolumn{2}{c|}{\textbf{Enron}} 
        & \multicolumn{3}{c|}{\textbf{ECHR}}  
        & \multirow{2}{*}{\shortstack{\textbf{Overall} \\ \textbf{Score}}} \\  
        \cmidrule(lr){2-3} \cmidrule(lr){5-6} \cmidrule(lr){7-9}
        & \textbf{Norm.}
        & \textbf{Aug.}
        &  
        & \textbf{Zero-S}
        & \textbf{Five-S}
        & \textbf{Name}
        & \textbf{Date}
        & \textbf{Loc.}
        &  \\ 
        \midrule
\includegraphics[width=0.33cm]{images/logo/openai-logo.png} \textbf{GPT-4o} & 97.50 & 100.00 & 82.69 & 99.50 & 93.00 & 91.50 & 94.50 & 74.00 & \textbf{91.09} \\
\includegraphics[width=0.33cm]{images/logo/deepseek-logo.png} \textbf{DeepSeek-R1} & 59.64 & 100.00 & 87.42 & 97.50 & 82.50 & 83.50 & 91.50 & 69.50 & \textbf{84.69} \\
\includegraphics[width=0.33cm]{images/logo/claude-logo-light.png} \textbf{Claude-3.5} & 83.21 & 100.00 & 78.69 & 76.00 & 50.00 & 93.00 & 94.50 & 82.00 & \textbf{80.78} \\
\includegraphics[width=0.33cm]{images/logo/claude-logo-light.png} \textbf{Claude-4.5} & 98.57 & 100.00 & 85.72 & 91.50 & 2.00 & 84.00 & 95.00 & 70.00 & \textbf{78.69} \\
\includegraphics[width=0.33cm]{images/logo/openai-logo.png} \textbf{GPT-5} & 94.93 & 100.00 & 81.42 & 11.50 & 8.54 & 93.00 & 94.50 & 78.00 & \textbf{69.36} \\
\includegraphics[width=0.33cm]{images/logo/grok-logo-light.png} \textbf{Grok 4} & 95.36 & 100.00 & 83.15 & 10.50 & 1.00 & 91.00 & 94.50 & 75.50 & \textbf{68.40} \\
\includegraphics[width=0.33cm]{images/logo/grok-logo-light.png} \textbf{Grok 4.1} & 93.21 & 100.00 & 84.04 & 7.00 & 1.00 & 92.00 & 95.00 & 77.00 & \textbf{68.16} \\
\includegraphics[width=0.33cm]{images/logo/grok-logo-light.png} \textbf{Grok 3} & 95.71 & 100.00 & 85.57 & 7.50 & 0.50 & 83.00 & 93.00 & 70.00 & \textbf{67.36} \\
\includegraphics[width=0.33cm]{images/logo/gemini-logo-light.png} \textbf{Gemini 3} & 87.86 & 100.00 & 84.53 & 4.00 & 0.00 & 84.00 & 91.00 & 65.00 & \textbf{65.12} \\
        \midrule
\includegraphics[width=0.33cm]{images/logo/huggingface-logo.png} \textbf{Phi-4} & 98.21 & 100.00 & 83.97 & 96.50 & 32.00 & 90.00 & 98.00 & 81.00 & \textbf{84.25} \\
\includegraphics[width=0.33cm]{images/logo/huggingface-logo.png} \textbf{GPT-120B} & 100.00 & 100.00 & 79.78 & 31.00 & 28.00 & 98.50 & 99.50 & 96.50 & \textbf{76.86} \\
\includegraphics[width=0.33cm]{images/logo/huggingface-logo.png} \textbf{Llama3.3-70B} & 89.64 & 100.00 & 80.89 & 40.50 & 29.00 & 89.50 & 95.00 & 76.50 & \textbf{74.36} \\
\includegraphics[width=0.33cm]{images/logo/huggingface-logo.png} \textbf{GPT-20B} & 98.93 & 100.00 & 72.65 & 28.50 & 7.50 & 97.50 & 100.00 & 96.00 & \textbf{71.99} \\
\includegraphics[width=0.33cm]{images/logo/huggingface-logo.png} \textbf{Gemma4-26B-A4B} & 93.93 & 100.00 & 85.28 & 21.50 & 0.50 & 91.00 & 95.00 & 83.50 & \textbf{70.77} \\
\includegraphics[width=0.33cm]{images/logo/huggingface-logo.png} \textbf{Gemma4-31B} & 88.21 & 100.00 & 84.11 & 21.50 & 0.50 & 92.50 & 94.00 & 81.00 & \textbf{69.60} \\
\includegraphics[width=0.33cm]{images/logo/huggingface-logo.png} \textbf{Qwen3.5-122B-A10B} & 72.14 & 77.50 & 57.65 & 11.50 & 0.00 & 81.00 & 93.00 & 60.00 & \textbf{54.06} \\
\includegraphics[width=0.33cm]{images/logo/huggingface-logo.png} \textbf{Qwen3.5-35B-A3B} & 48.93 & 62.14 & 66.90 & 27.00 & 0.50 & 81.50 & 93.50 & 64.00 & \textbf{53.96} \\
\includegraphics[width=0.33cm]{images/logo/huggingface-logo.png} \textbf{Qwen3.5-9B} & 34.64 & 89.29 & 60.07 & 12.00 & 2.00 & 82.50 & 93.00 & 69.50 & \textbf{52.68} \\
\includegraphics[width=0.33cm]{images/logo/huggingface-logo.png} \textbf{Qwen3.5-4B} & 29.29 & 61.07 & 67.94 & 14.50 & 2.00 & 84.00 & 92.50 & 72.00 & \textbf{51.05} \\
\includegraphics[width=0.33cm]{images/logo/huggingface-logo.png} \textbf{Qwen3.5-27B} & 26.43 & 68.93 & 55.23 & 23.50 & 0.50 & 82.00 & 90.00 & 63.50 & \textbf{48.35} \\

        \bottomrule
    \end{tabular}
    \end{adjustbox}
    \label{tab:data_privacy_results}
\end{table}

\begin{table}[!t]
    \centering
    \caption{\system Fairness \& Bias Leaderboard}
    \renewcommand{\arraystretch}{1.3} 
    \resizebox{\linewidth}{!}{
    \begin{tabular}{l|ccccc|ccc|cc|c|c}
        \toprule
        \multirow{2}{*}{\textbf{Model}} & \multicolumn{5}{c|}{\textbf{Adult}} & \multicolumn{3}{c|}{\textbf{Gendercare}} & \multicolumn{2}{c|}{\textbf{Preference}} & \textbf{BBQ} & \multirow{2}{*}{\shortstack{\textbf{Overall} \\ \textbf{Score}}} \\
        \cmidrule(lr){2-6} \cmidrule(lr){7-9} \cmidrule(lr){10-11}
        & \textbf{Sex} 
        & \textbf{Race} 
        & \textbf{Edu} 
        & \textbf{Hours} 
        & \textbf{Type} 
        & \textbf{M} 
        & \textbf{F} 
        & \textbf{N} 
        & \rotatebox{75}{\textbf{Lifestyle}} 
        & \rotatebox{75}{\textbf{Ideology}} 
        & & \\
        \midrule
\includegraphics[width=0.33cm]{images/logo/gemini-logo-light.png} \textbf{Gemini 3} & 90.11 & 92.21 & 71.56 & 92.52 & 58.04 & 74.07 & 74.01 & 73.87 & 4.88 & 86.08 & 92.18 & \textbf{76.72} \\
\includegraphics[width=0.33cm]{images/logo/openai-logo.png} \textbf{GPT-5} & 92.22 & 92.45 & 72.92 & 91.52 & 52.82 & 82.09 & 73.36 & 77.16 & 0.00 & 34.81 & 92.41 & \textbf{68.78} \\
\includegraphics[width=0.33cm]{images/logo/openai-logo.png} \textbf{GPT-4o} & 87.09 & 94.22 & 72.54 & 92.75 & 44.67 & 75.78 & 69.71 & 72.13 & 7.32 & 32.91 & 93.55 & \textbf{67.53} \\
\includegraphics[width=0.33cm]{images/logo/claude-logo-light.png} \textbf{Claude-4.5} & 91.31 & 94.88 & 65.75 & 96.93 & 41.67 & 71.72 & 68.52 & 71.36 & 0.00 & 25.95 & 83.36 & \textbf{62.71} \\
\includegraphics[width=0.33cm]{images/logo/deepseek-logo.png} \textbf{DeepSeek-R1} & 93.90 & 95.07 & 65.04 & 96.17 & 39.42 & 73.06 & 70.82 & 65.83 & 0.00 & 5.06 & 93.09 & \textbf{61.59} \\
\includegraphics[width=0.33cm]{images/logo/grok-logo-light.png} \textbf{Grok 4} & 92.10 & 93.71 & 74.82 & 94.13 & 34.94 & 69.36 & 68.85 & 74.87 & 0.00 & 0.00 & 94.18 & \textbf{61.50} \\
\includegraphics[width=0.33cm]{images/logo/claude-logo-light.png} \textbf{Claude-3.5} & 94.04 & 93.11 & 61.15 & 95.67 & 43.52 & 74.41 & 69.18 & 65.83 & 0.00 & 16.46 & 85.55 & \textbf{61.45} \\
\includegraphics[width=0.33cm]{images/logo/grok-logo-light.png} \textbf{Grok 3} & 80.57 & 94.59 & 67.85 & 96.04 & 49.72 & 74.07 & 69.51 & 67.59 & 0.00 & 0.00 & 94.45 & \textbf{60.92} \\
\includegraphics[width=0.33cm]{images/logo/grok-logo-light.png} \textbf{Grok 4.1} & 88.78 & 94.03 & 75.02 & 93.45 & 36.48 & 62.29 & 65.57 & 66.33 & 0.00 & 0.00 & 93.00 & \textbf{59.53} \\
        \midrule
\includegraphics[width=0.33cm]{images/logo/huggingface-logo.png} \textbf{Gemma4-31B} & 96.62 & 96.99 & 78.33 & 94.18 & 35.96 & 76.09 & 74.75 & 75.13 & 8.54 & 92.41 & 94.82 & \textbf{78.96} \\
\includegraphics[width=0.33cm]{images/logo/huggingface-logo.png} \textbf{Gemma4-26B-A4B} & 95.64 & 92.18 & 63.97 & 94.27 & 45.78 & 73.74 & 74.75 & 72.11 & 10.98 & 91.14 & 92.64 & \textbf{77.50} \\
\includegraphics[width=0.33cm]{images/logo/huggingface-logo.png} \textbf{GPT-120B} & 94.53 & 93.59 & 72.97 & 95.69 & 35.54 & 74.07 & 72.46 & 70.85 & 10.98 & 62.66 & 92.82 & \textbf{72.93} \\
\includegraphics[width=0.33cm]{images/logo/huggingface-logo.png} \textbf{Phi-4} & 92.04 & 89.70 & 73.82 & 98.76 & 43.31 & 65.99 & 66.23 & 62.81 & 31.71 & 60.13 & 92.55 & \textbf{72.01} \\
\includegraphics[width=0.33cm]{images/logo/huggingface-logo.png} \textbf{Qwen3.5-27B} & 94.61 & 97.46 & 85.85 & 94.92 & 87.81 & 54.21 & 55.41 & 39.45 & 13.41 & 77.22 & 88.00 & \textbf{71.17} \\
\includegraphics[width=0.33cm]{images/logo/huggingface-logo.png} \textbf{GPT-20B} & 95.02 & 92.30 & 69.62 & 91.27 & 56.05 & 73.74 & 70.49 & 67.34 & 7.32 & 38.61 & 90.00 & \textbf{67.64} \\
\includegraphics[width=0.33cm]{images/logo/huggingface-logo.png} \textbf{Qwen3.5-122B-A10B} & 98.65 & 95.18 & 87.25 & 90.36 & 77.87 & 35.02 & 42.95 & 22.36 & 14.63 & 63.92 & 88.00 & \textbf{64.32} \\
\includegraphics[width=0.33cm]{images/logo/huggingface-logo.png} \textbf{Qwen3.5-9B} & 88.38 & 82.84 & 88.88 & 95.79 & 91.56 & 23.57 & 25.90 & 11.56 & 23.17 & 77.85 & 88.45 & \textbf{63.78} \\
\includegraphics[width=0.33cm]{images/logo/huggingface-logo.png} \textbf{Qwen3.5-35B-A3B} & 96.46 & 86.40 & 92.53 & 92.14 & 67.77 & 25.25 & 21.64 & 13.32 & 21.95 & 76.58 & 85.82 & \textbf{63.03} \\
\includegraphics[width=0.33cm]{images/logo/huggingface-logo.png} \textbf{Qwen3.5-4B} & 82.34 & 68.17 & 76.77 & 93.29 & 88.91 & 18.18 & 16.39 & 9.30 & 21.95 & 65.82 & 88.73 & \textbf{59.29} \\
\includegraphics[width=0.33cm]{images/logo/huggingface-logo.png} \textbf{Llama3.3-70B} & 87.71 & 94.27 & 65.14 & 93.25 & 43.07 & 71.38 & 66.89 & 68.09 & 0.00 & 1.27 & 87.27 & \textbf{58.92} \\

        \bottomrule
    \end{tabular}
    }
    \label{tab:fairness_bias_results}
\end{table}

\subsection{Model \& Data Privacy}
\label{res:privacy}

Table~\ref{tab:data_privacy_results} shows that both proprietary and open-weight models can achieve strong privacy, though with distinct patterns across test categories. GPT-4o leads among API models, while Phi-4 achieves the highest score among open-weight models. Most models benefit substantially from explicit privacy reminders, indicating strong reliance on prompt-level privacy guidance.

\textit{ConfAIDe} evaluations reveal a clearer gap: proprietary models align more closely with human privacy expectations, while many open-weight models, particularly smaller variants, exhibit more variability. On \textit{Enron}, a stark divide emerges: older models such as GPT-4o and DeepSeek-R1 effectively resist email-address extraction, whereas several newer models (GPT-5, Grok variants, and Gemini~3) exhibit dramatically higher leakage in the zero-shot category, a pattern that worsens in the five-shot category. \textit{ECHR} results further show that models are generally reliable with name and date fields but often struggle with location, reflecting a persistent weakness in handling geographic PII.

\subsection{Fairness \& Bias}\label{res:bias}

Table~\ref{tab:fairness_bias_results} shows that models generally achieve high parity on the \textit{Adult} benchmark, indicating stable performance across core demographic attributes. However, performance drops on \textit{GenderCARE}, where disparities in gendered word-choice selection remain pronounced even among top models. The \textit{Preference} test remains the most challenging: nearly all models struggle to maintain ideological and lifestyle neutrality, rarely refusing to state a preference. In contrast, \textit{BBQ} reveals strong stereotype-avoidance in several models.

\end{document}